\documentclass[fleqn,usenatbib]{mnras}

\usepackage{newtxtext,newtxmath}

\usepackage[T1]{fontenc}

\DeclareRobustCommand{\VAN}[3]{#2}
\let\VANthebibliography\thebibliography
\def\thebibliography{\DeclareRobustCommand{\VAN}[3]{##3}\VANthebibliography}

\usepackage{graphicx}
\usepackage{multirow}
\usepackage{graphicx}	
\usepackage{amsmath}	
\usepackage{placeins}
\usepackage{xfrac}

\usepackage{hyperref}
\hypersetup{colorlinks=true, linkcolor=Navy, citecolor=Navy, urlcolor=Navy}

\usepackage{color}

\definecolor{DarkRed}           {RGB}{139,   0,   0}
\definecolor{Red}               {RGB}{255,   0,   0}
\definecolor{Firebrick}         {RGB}{178,  34,  34}
\definecolor{Crimson}           {RGB}{220,  20,  60}
\definecolor{IndianRed}         {RGB}{205,  92,  92}
\definecolor{LightCoral}        {RGB}{240, 128, 128}
\definecolor{Salmon}            {RGB}{250, 128, 114}
\definecolor{DarkSalmon}        {RGB}{233, 150, 122}
\definecolor{LightSalmon}       {RGB}{255, 160, 122}

\definecolor{OrangeRed}         {RGB}{255,  69,   0}
\definecolor{Tomato}            {RGB}{255,  99,  71}
\definecolor{DarkOrange}        {RGB}{255, 140,   0}
\definecolor{Coral}             {RGB}{255, 127,  80}
\definecolor{Orange}            {RGB}{255, 165,   0}

\definecolor{DarkKhaki}         {RGB}{189, 183, 107}
\definecolor{Gold}              {RGB}{255, 215,   0}
\definecolor{Khaki}             {RGB}{240, 230, 140}
\definecolor{PeachPuff}         {RGB}{255, 218, 185}
\definecolor{Yellow}            {RGB}{255, 255,   0}
\definecolor{PaleGoldenrod}     {RGB}{238, 232, 170}
\definecolor{Moccasin}          {RGB}{255, 228, 181}
\definecolor{PapayaWhip}        {RGB}{255, 239, 213}
\definecolor{LightGoldenrodYellow}{RGB}{250, 250, 210}
\definecolor{LemonChiffon}      {RGB}{255, 250, 205}
\definecolor{LightYellow}       {RGB}{255, 255, 224}

\definecolor{Maroon}            {RGB}{128,   0,   0}
\definecolor{Brown}             {RGB}{165,  42,  42}
\definecolor{SaddleBrown}       {RGB}{139,  69,  19}
\definecolor{Sienna}            {RGB}{160,  82,  45}
\definecolor{Chocolate}         {RGB}{210, 105,  30}
\definecolor{DarkGoldenrod}     {RGB}{184, 134,  11}
\definecolor{Peru}              {RGB}{205, 133,  63}
\definecolor{RosyBrown}         {RGB}{188, 143, 143}
\definecolor{Goldenrod}         {RGB}{218, 165,  32}
\definecolor{SandyBrown}        {RGB}{244, 164,  96}
\definecolor{Tan}               {RGB}{210, 180, 140}
\definecolor{Burlywood}         {RGB}{222, 184, 135}
\definecolor{Wheat}             {RGB}{245, 222, 179}
\definecolor{NavajoWhite}       {RGB}{255, 222, 173}
\definecolor{Bisque}            {RGB}{255, 228, 196}
\definecolor{BlanchedAlmond}    {RGB}{255, 235, 205}
\definecolor{Cornsilk}          {RGB}{255, 248, 220}

\definecolor{DarkGreen}         {RGB}{  0, 100,   0}
\definecolor{Green}             {RGB}{  0, 128,   0}
\definecolor{DarkOliveGreen}    {RGB}{ 85, 107,  47}
\definecolor{ForestGreen}       {RGB}{ 34, 139,  34}
\definecolor{SeaGreen}          {RGB}{ 46, 139,  87}
\definecolor{Olive}             {RGB}{128, 128,   0}
\definecolor{OliveDrab}         {RGB}{107, 142,  35}
\definecolor{MediumSeaGreen}    {RGB}{ 60, 179, 113}
\definecolor{LimeGreen}         {RGB}{ 50, 205,  50}
\definecolor{Lime}              {RGB}{  0, 255,   0}
\definecolor{SpringGreen}       {RGB}{  0, 255, 127}
\definecolor{MediumSpringGreen} {RGB}{  0, 250, 154}
\definecolor{DarkSeaGreen}      {RGB}{143, 188, 143}
\definecolor{MediumAquamarine}  {RGB}{102, 205, 170}
\definecolor{YellowGreen}       {RGB}{154, 205,  50}
\definecolor{LawnGreen}         {RGB}{124, 252,   0}
\definecolor{Chartreuse}        {RGB}{127, 255,   0}
\definecolor{LightGreen}        {RGB}{144, 238, 144}
\definecolor{GreenYellow}       {RGB}{173, 255,  47}
\definecolor{PaleGreen}         {RGB}{152, 251, 152}

\definecolor{Teal}              {RGB}{  0, 128, 128}
\definecolor{DarkCyan}          {RGB}{  0, 139, 139}
\definecolor{LightSeaGreen}     {RGB}{ 32, 178, 170}
\definecolor{CadetBlue}         {RGB}{ 95, 158, 160}
\definecolor{DarkTurquoise}     {RGB}{  0, 206, 209}
\definecolor{MediumTurquoise}   {RGB}{ 72, 209, 204}
\definecolor{Turquoise}         {RGB}{ 64, 224, 208}
\definecolor{Aqua}              {RGB}{  0, 255, 255}
\definecolor{Cyan}              {RGB}{  0, 255, 255}
\definecolor{Aquamarine}        {RGB}{127, 255, 212}
\definecolor{PaleTurquoise}     {RGB}{175, 238, 238}
\definecolor{LightCyan}         {RGB}{224, 255, 255}

\definecolor{Navy}              {RGB}{  0,   0, 128}
\definecolor{DarkBlue}          {RGB}{  0,   0, 139}
\definecolor{MediumBlue}        {RGB}{  0,   0, 205}
\definecolor{Blue}              {RGB}{  0,   0, 255}
\definecolor{MidnightBlue}      {RGB}{ 25,  25, 112}
\definecolor{RoyalBlue}         {RGB}{ 65, 105, 225}
\definecolor{SteelBlue}         {RGB}{ 70, 130, 180}
\definecolor{DodgerBlue}        {RGB}{ 30, 144, 255}
\definecolor{DeepSkyBlue}       {RGB}{  0, 191, 255}
\definecolor{CornflowerBlue}    {RGB}{100, 149, 237}
\definecolor{SkyBlue}           {RGB}{135, 206, 235}
\definecolor{LightSkyBlue}      {RGB}{135, 206, 250}
\definecolor{LightSteelBlue}    {RGB}{176, 196, 222}
\definecolor{LightBlue}         {RGB}{173, 216, 230}
\definecolor{PowderBlue}        {RGB}{176, 224, 230}

\definecolor{Indigo}            {RGB}{ 75,   0, 130}
\definecolor{Purple}            {RGB}{128,   0, 128}
\definecolor{DarkMagenta}       {RGB}{139,   0, 139}
\definecolor{DarkViolet}        {RGB}{148,   0, 211}
\definecolor{DarkSlateBlue}     {RGB}{ 72,  61, 139}
\definecolor{BlueViolet}        {RGB}{138,  43, 226}
\definecolor{DarkOrchid}        {RGB}{153,  50, 204}
\definecolor{Fuchsia}           {RGB}{255,   0, 255}
\definecolor{Magenta}           {RGB}{255,   0, 255}
\definecolor{SlateBlue}         {RGB}{106,  90, 205}
\definecolor{MediumSlateBlue}   {RGB}{123, 104, 238}
\definecolor{MediumOrchid}      {RGB}{186,  85, 211}
\definecolor{MediumPurple}      {RGB}{147, 112, 219}

\definecolor{MistyRose}         {RGB}{255, 228, 225}
\definecolor{AntiqueWhite}      {RGB}{250, 235, 215}
\definecolor{Linen}             {RGB}{250, 240, 230}
\definecolor{Beige}             {RGB}{245, 245, 220}
\definecolor{WhiteSmoke}        {RGB}{245, 245, 245}
\definecolor{LavenderBlush}     {RGB}{255, 240, 245}
\definecolor{OldLace}           {RGB}{253, 245, 230}
\definecolor{AliceBlue}         {RGB}{240, 248, 255}
\definecolor{Seashell}          {RGB}{255, 245, 238}
\definecolor{GhostWhite}        {RGB}{248, 248, 255}
\definecolor{Honeydew}          {RGB}{240, 255, 240}
\definecolor{FloralWhite}       {RGB}{255, 250, 240}
\definecolor{Azure}             {RGB}{240, 255, 255}
\definecolor{MintCream}         {RGB}{245, 255, 250}
\definecolor{Snow}              {RGB}{255, 250, 250}
\definecolor{Ivory}             {RGB}{255, 255, 240}
\definecolor{White}             {RGB}{255, 255, 255}

\definecolor{Black}             {RGB}{  0,   0,   0}
\definecolor{DarkSlateGray}     {RGB}{ 47,  79,  79}
\definecolor{DimGray}           {RGB}{105, 105, 105}
\definecolor{SlateGray}         {RGB}{112, 128, 144}
\definecolor{Gray}              {RGB}{128, 128, 128}
\definecolor{LightSlateGray}    {RGB}{119, 136, 153}
\definecolor{DarkGray}          {RGB}{169, 169, 169}
\definecolor{Silver}            {RGB}{192, 192, 192}
\definecolor{LightGray}         {RGB}{211, 211, 211}
\definecolor{Gainsboro}         {RGB}{220, 220, 220}

\definecolor{DavysGray}         {RGB}{ 85,  85,  85}
\definecolor{Jet}               {RGB}{ 52,  52,  52}

\newcommand{\AAA}{\textup{\AA}}
\newcommand{\kmps}{\textup{km/s}}
\newcommand{\mps}{\textup{m/s}}
\newcommand{\cmps}{\textup{cm/s}}

\newcommand{\hl}[1]{#1}

\title[The ESPRESSO Line-Spread Function]{Characterization of the ESPRESSO Line-Spread Function and Improvement of the Wavelength Calibration Accuracy}

\author[T. M. Schmidt]{
Tobias M. Schmidt,$^{1}$\thanks{E-mail: tobias.schmidt@unige.ch}, Fran\c{c}ois Bouchy$^{1}$\\
$^{1}$Observatoire Astronomique de l'Universit\'e de Gen\`eve, Chemin Pegasi 51, Sauverny, CH-1290, Switzerland
}
\date{Accepted XXX. Received YYY; in original form ZZZ}

\pubyear{2024}

\begin{document}
\graphicspath{{./}}
\label{firstpage}
\pagerange{\pageref{firstpage}--\pageref{lastpage}}
\maketitle

\begin{abstract}
Achieving a truly accurate wavelength calibration of high-dispersion echelle spectrographs is a challenging task but crucially needed for certain science cases, e.g. to test for a possible variation of the fine-structure constant in quasar spectra.
One of the spectrographs best suited for this mission is VLT/ESPRESSO. Nevertheless, previous studies have identified significant discrepancies between the classical wavelength solutions and the one derived independently from the laser frequency comb. The dominant parts of these systematics were intra-order distortions, most-likely related to a deviation of the instrumental line-spread function from the assumed Gaussian shape.
Here, we therefore present a study focused on a detailed modeling of the ESPRESSO instrumental line-spread function.
We demonstrate that it is strongly asymmetric, non-Gaussian, different for the two slices and fibers, and varies significantly along the spectral orders.
Incorporating the determined non-parametric model in the wavelength calibration process drastically improves the wavelength calibration accuracy, reducing the discrepancies between the two independent wavelength solutions from $50\,\mps$ to about $10\,\mps$.
The most striking success is, however, that the different fibers and slices now provide fully consistent measurements with a scatter of just a couple $\mps$. This demonstrates that the instrument-related systematics can be nearly eliminated over most of the spectral range by properly taking into account the complex shape of the instrumental line-spread function and paves the way for further optimizations of the wavelength calibration process.

\end{abstract}



\section{Introduction}

The Echelle SPectrograph for Rocky Exoplanets and Stable Spectroscopic Observations (ESPRESSO, \citealt{Pepe2021}), installed since 2018 at the ESO VLT observatory, is the current flagship for stable and accurate spectroscopic observations.
Two of its main science drivers are the hunt for extrasolar planets using the radial-velocity (RV) method, pioneered by \citet{Mayor1995}, and the search for a possible change of fundamental physical constants \cite[e.g.][]{Webb1999, King2012, Murphy2022}.
\hl{In particular these two science cases crucially rely on a high-quality wavelength calibration of the spectrograph, but their specific requirement are in fact of rather different nature, one requiring \textit{stability}, the other \textit{accuracy} of the wavelength calibration.}
\hl{
To clarify this aspect, a detailed description of the adopted definition for \textit{precision}, \textit{accuracy}, and \textit{stability}, is given in Appendix~\ref{Sec:PAS}.%
}
\hl{In short}, RV searches require \textit{stability} of the wavelength calibration over the course of the series of observations, which might take weeks to years. The goal for ESPRESSO is a RV stability (in the final data products and after all calibrations have been applied) of $10\,\cmps$, sufficient for the detection of a twin-Earth.
The search for a varying fine-structure constant, however, imposes a totally different set of requirements. Here, \textit{accuracy} is needed, since the value of the fine-structure constant is inferred by the accurate wavelength measurement of different metal absorption lines in the same quasar spectrum. Therefore, the wavelength scale actually needs to be correct and, most crucially, free of distortions.
To reliably detect a relative change of the fine-structure constant by 1\,ppm, such distortions must be below $\approx 20\,\mps$ (peak-to-valley).
\hl{A more extensive description of the measurement principle used to infer the fine-structure constant from quasar spectra and the implied requirements is given in Appendix~\ref{FineStructureConstant}.}

To pave the way for the fine-structure studies conducted as part of the ESPRESSO GTO program \citep[][]{Murphy2022b}, a thorough assessment and analysis of the ESPRESSO wavelength calibration accuracy has been presented in \citet{Schmidt2021}.
The study provided vital insights into the instrument and confirmed the overall good quality of the wavelength solution. However, it also revealed numerous instrument systematics and inaccuracies. Improvements to the wavelength calibration accuracy are thus needed, also in view of ANDES \citep[ArmazoNes high Dispersion Echelle Spectrograph,][]{Marconi2022}, the future high-resolution spectrograph for the ELT \citep{Padovani2023}, which shall continue the work of ESPRESSO and, thanks to the much larger photon-collecting area of the ELT, push current limit by about a factor of five \citep{Martins2023}. This, however, will only be possible if also the accuracy of the wavelength calibration is improved in a similar manner. Here, ESPRESSO acts as an excellent testbed to develop novel methods and algorithms for wavelength calibration and to demonstrate that wavelength calibration at the quality level required for ANDES is actually possible.

In this study, we therefore continue our initial work presented in \citet{Schmidt2021} and develop novel methods to significantly improve the ESPRESSO wavelength calibration accuracy. One of the most noticeable instrumental systematics found by \citet{Schmidt2021} were intra-order wavelength distortions. It was quickly suspected that these might relate to the instrumental line-spread function (LSF), which probably is non-Gaussian, asymmetric, and might vary significantly along individual spectral orders. This could easily introduce the observed systematics when not properly accounting for it. Also, deviations of the LSF from a simple Gaussian shape are expected for high-resolution echelle spectrographs and have been reported for other instruments by numerous authors \citep[e.g][]{Butler1996, Kambe2002, Sato2002, Zhao2014, Hirano2020, Zhao2021, Terrien2021, Milakovic2024}.
Still, there are few astronomical spectrographs for which the shape of the instrumental LSF is properly characterized.
The focus of this study is therefore to accurately measure the ESPRESSO LSF, construct a model of it, and incorporate this knowledge in the wavelength calibration process, producing a substantially more accurate wavelength calibration.

\section{Instrumental Setup}

ESPRESSO is a fiber-fed high-resolution spectrograph installed at the ESO Paranal observatory \citep{Molaro2009, Pepe2010, Pepe2014, Pepe2021}.
Like many other RV spectrographs, ESPRESSO can be simultaneously fed by two fibers, one to collect the light from an astronomical target, while the other receives light from a simultaneous wavelength calibration source (or the sky). For calibration purposes, one can also \hl{just} feed light from one or two different calibration sources to the spectrograph, which is used within the context of this study. To decrease the physical size of the spectrograph, the ESPRESSO optical design incorporates a pupil slicer \citep{Riva2014a}. Therefore, the spectra corresponding to each of the two fibers are imaged twice onto the detectors. In the default high-resolution mode, these can be extracted separately, providing formally independent spectra of the same source, and allow valuable cross-checks between slices and fibers.

Extremely important for test and verification of the wavelength calibration is furthermore that
ESPRESSO is equipped with multiple calibration sources that permit to derive two fully independent wavelength solutions.
The default wavelength calibration is based on a combination of a thorium-argon hollow-cathode lamp (ThAr HCL) and a Fabry-P\'erot interferometer \citep[FP, ][]{Wildi2010, Wildi2011, Wildi2012}. Here, the ThAr HCL provides absolute wavelength information and is used to calibrate the FP, while the passively-stabilized FP offers a densely-spaced ensemble of equally-bright lines covering the full wavelength range and allows to construct a high-fidelity wavelength solution, basically an interpolation of the ThAr lines in wavelength and time \citep{Cersullo2019, Schmidt2021}.

In addition to that, ESPRESSO is equipped with a laser frequency comb (LFC).
This is a special kind of laser device that produces a dense, broadband spectrum composed of many narrow lines equally-spaced in frequency, called a \textit{comb}. Key feature is that the frequencies of those lines are measured and actively controlled to extreme accuracy, linking them directly to the fundamental SI time standard, and enabling unprecedented metrological measurements \citep{Reichert1999, Jones2000, Udem2002}.
The ESPRESSO LFC is based on a $1\,\mu\textup{m}$ mode-locked femtosecond laser with a repetition rate of $250\,\textup{MHz}$, filtered by Fabry-P\'erot cavities to a line separation of $18\,\textup{GHz}$, and broadened into the optical regime using a tapered photonic crystal fiber (PCF) \citep[see e.g.][for a description of a similar LFC system]{Probst2016}.
It allows to construct a separate, fully independent wavelength solution. In addition, the LFC spectra, composed of a plethora of extremely narrow lines, directly reveal the instrumental LSF, which is crucial for our endeavor to characterize and incorporate the LSF in the wavelength calibration process.
Only the availability of those features, i.e. two independent sets of wavelength calibrations, a source providing truly unresolved lines, and a spectrograph design with two fibers and slices, allows the accurate calibration and thorough testing of the ESPRESSO wavelength calibration we present in the following.

For this, we make use of the standard calibration exposures acquired in \texttt{1HR2x1} mode on January 24, 2023, as part of the daily calibration scheme.
All data was reduced and processed, starting from the raw frames, using a custom data reduction and calibration pipeline.
The general concept was outlined in \citet{Schmidt2021} and in its basic functionality, our pipeline follows the same strategy and uses similar algorithms as the official ESPRESSO DRS%
\footnote{\url{https://www.eso.org/sci/software/pipelines/index.html}}. %
However, over the last years, it has been continuously improved, of which the biggest change is the LSF determination and inclusion thereof in the wavelength calibration, described in this paper.

\section{LFC Properties after 2022 Upgrades}
\label{Sec:LFC_Flux}

\begin{figure*}
 \includegraphics[width=\linewidth]{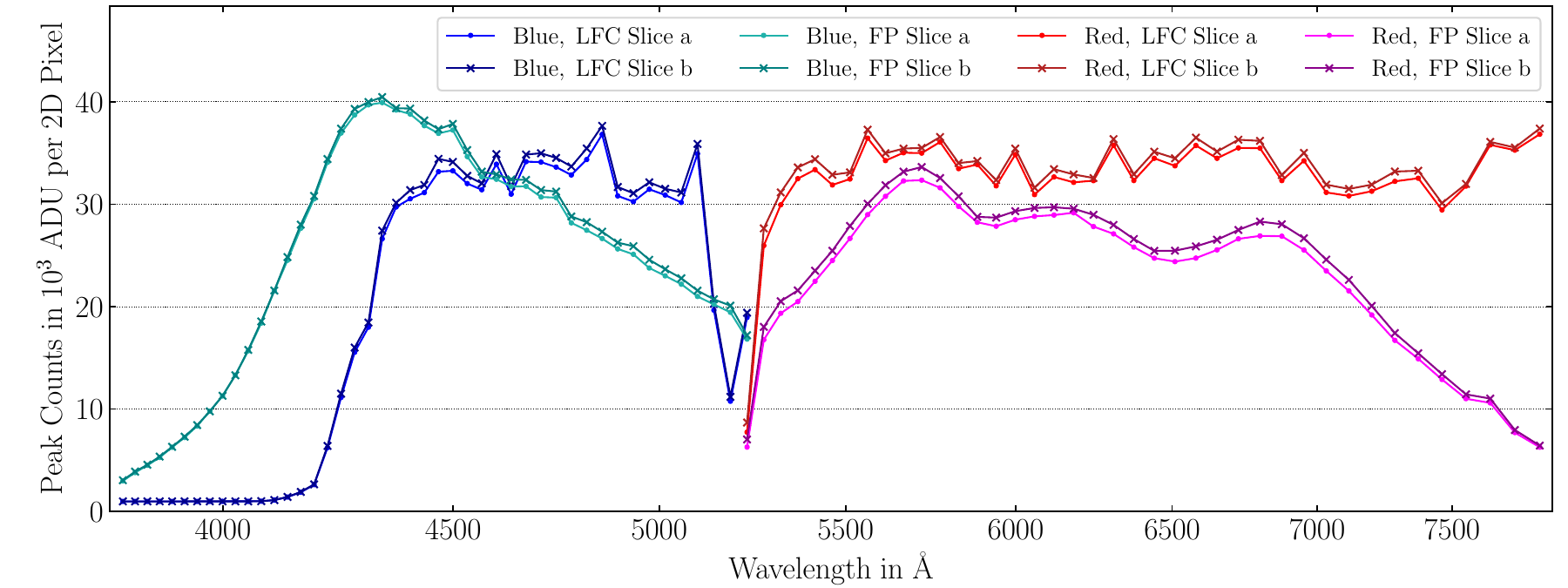}
 \caption{
  Flux levels of LFC (Fiber~A) and FP (Fiber~B) spectra in a standard \texttt{WAVE,LFC,FP} wavelength calibration frame, taken on January~24, 2023, as part of the daily calibration plan. The shown quantity are the peak counts per binned  2D pixel (in \texttt{1HR2x1} mode) for each spectral order of ESPRESSO.
 }
 \label{Fig:Flux_LCFP}
\end{figure*}

In May~2022, significant upgrades to the ESPRESSO LFC system have been performed. Several components were exchanged, the thermal stability of the system improved and, in particular, a new PCF installed, which is responsible for the spectral broadening of the near-IR comb into the visible regime. A further intervention in October and November 2022 optimized the system, mostly in terms of spectral flatness and flux level, and brought it back into operation. Since then, daily calibrations are taken as part of the standard instrument calibration plan and are publicly available from the ESO archive.

The new PCF offers substantially increased wavelength coverage, now reaching down to $\approx4300\,\mathrm{\AAA}$, much farther than the $\approx4900\,\mathrm{\AAA}$ that were available until November 2019, when the LFC ceased operation. Upgrades to other components now also allow to cover the reddest part ($\lambda>6900\,\mathrm{\AAA}$) of the ESPRESSO spectral range, together increasing the coverage from approx $56\%$ to $82\%$ of the spectrograph's spectral range.
The disadvantage of the new PCF is that it provides much less flux. To compensate, the liquid core fiber was removed from the active mode scrambler \citep{Frank2018}, which increased the throughput by about one order of magnitude.  In addition, the exposure time of the \texttt{WAVE,LFC,FP} and \texttt{WAVE,FP,LFC} calibration frames had to be increased to $180\,\mathrm{s}$. 
Figure~\ref{Fig:Flux_LCFP} illustrates the flux level of a standard \texttt{WAVE,LFC,FP} frame taken in \texttt{1HR2x1} mode. For each spectral order, the peak counts in ADU per binned 2D pixel are given. The flux distribution within each order of course follows the spectrograph's blaze function. Since the detector becomes non-linear and saturates at around 60k ADU, the desired peak exposure level should be high but safely below this limit.
The spectral flattening unit of the LFC was tuned to achieve a very flat peak exposure level of approx 32k~ADU for $\lambda>4400\,\mathrm{\AAA}$. The exception is a narrow spectral region around $5150\,\mathrm{\AAA}$ where the LFC provides far less flux. A completely flat spectrum could probably have been achieved by further attenuating the spectrum in the other regions. However, since the LFC flux levels were already low, it was decided to accept low flux in this region but get high flux everywhere else.
Throughout the year 2023, the LFC flux gradually decreased, to a level of barely 5k~ADU (peak) in Fall 2023. An intervention to the system in November 2023 brought this back to a reasonable level.

A re-occuring issue is that a certain fraction of LFC frames turn out to be unusable. Here, parts of the red detector are flooded with light, amounts that would correspond to peak exposure levels of 300k~ADU, driving the detector deep into saturation.
It appears that in such cases the spectral flattening unit--which is re-configured before each LFC exposure to ensure an optimal spectrum--has not reached its desired state, delivering far too much flux in the red part of the spectrum.
This issue seems to occur randomly for about 10\% of the exposures.

Despite regular operations of the ESPRESSO LFC, there exist some doubts about its performance and it remains to be demonstrated that an actually superior wavelength calibration can be obtained from the LFC spectra. Here, within the context of this paper, we focus on the \textit{accuracy} aspect and how it can be improved utilizing a LFC-derived model of the instrumental LSF. A thorough assessment of the achievable \textit{stability} warrants its own dedicated study, which will be presented somewhere else.

\FloatBarrier
\section{ESPRESSO Wavelength Calibration Accuracy}

\begin{figure*}
 \includegraphics[width=\linewidth]{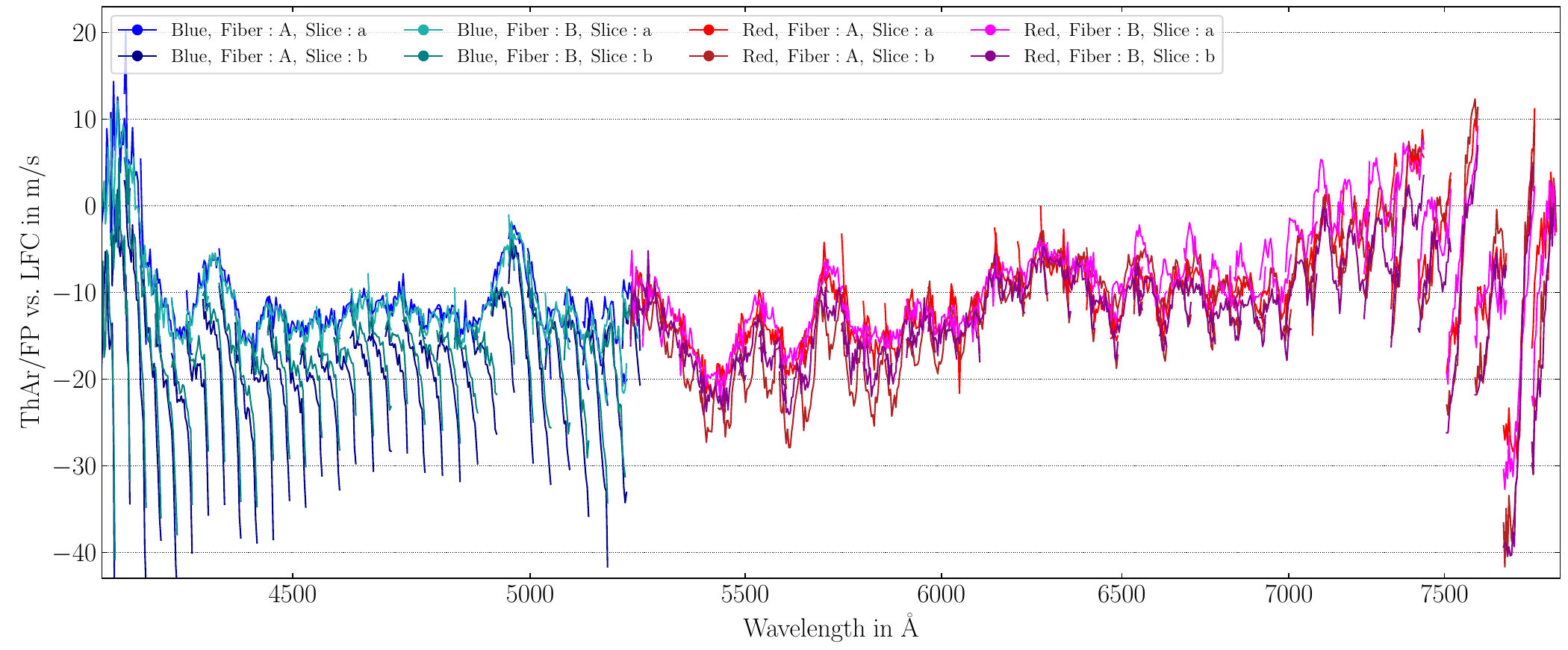}
 \caption{
  Comparison between the ThAr/FP and LFC wavelength solution, similar to the results presented in \citet{Schmidt2021}. However, the data shown here was taken on 2023-01-24 and covers, due to the LFC upgrade in Spring 2022, a much wider wavelength range. Therefore, the discrepancies between fibers and slices (indicated by various colors) and intra-order distortions become much more apparent.
 }
 \label{Fig:Comparison_ThArFP-LFC_Gauss}
\end{figure*}

In \citet{Schmidt2021}, we have presented a comparison of the combined ThAr/FP wavelength solution to the LFC-based one. Now, with the improved LFC system, we can perform the same comparison over a much wider wavelength range, which is shown in Figure~\ref{Fig:Comparison_ThArFP-LFC_Gauss}. Here, numerous systematics become a lot more apparent.
First of all, there is a global offset between the two wavelength solutions of approximately $-10\,\mps$, but also some variation on large spectral scales. In addition, there is modulation on small scales that correlates with the order structure of the spectrograph. For the red arm, one can notice increasing intra-order modulations towards the red end of the spectral range. For the five reddest spectral orders, the wavelength solutions in the overlap region of neighboring orders are not consistent at all anymore.
In the blue arm, one finds a strong discrepancy between slices. For both fibers, Slice~b exhibits a strong negative deviation in the ThAr/FP vs. LFC comparison towards the long-wavelength end of each spectral order, while Slice~a provides consistent results across orders. Interestingly, this behavior is quite similar for both fibers. All together, this results in peak-to-valley distortions%
\footnote{For a fine-structure constant measurement, a global offset is actually no problem, as long as it is constant in velocity for all wavelengths. Therefore, the peak-to-valley distortion across the spectral range is the relevant quantity.} %
of the wavelength calibration of over $40\,\mps$ and can therefore become the limiting factor for measurements of the fine-structure constant \citep[e.g.][]{Schmidt2021,Murphy2022}.
While the global offset and the large-scale structure might be related to the ThAr reference, either residual blends or inaccurate laboratory wavelengths, the discrepancy between fibers and slices is obviously related to ESPRESSO alone. Whatever input spectra are fed to the spectrograph, fully consistent results should be obtained from all fibers and slices. Similarly, the modulation on small scales strongly correlates with the spectral orders of the echelle spectrograph. This obviously is as well an artifact from the instrument and can not be related to the calibration sources either.

\hl{Admittedly, the presented comparison is not directly a demonstration of the absolute wavelength calibration accuracy. As described in detail in Appendix~\ref{Sec:PAS}, there exists no \textit{ground-truth} reference on sky that could be used to validate the wavelength calibration accuracy of a spectrograph and one thus has to always resort to partial or approximate validations schemes.
Nevertheless, the comparison of independent calibration sources (ThAr/FP vs. LFC) presents the best possible test that can be done with the equipment at hand and, due to the vastly different nature of the involved calibration sources, becomes quite representative for the true wavelength calibration accuracy of ESPRESSO. In particular}, it sets a clear lower limit to it: As long as these two fully independent, but by themselves completely viable%
\footnote{We have no reason to assume one of the two wavelength solutions to be \textit{better} than the other. In fact, they both come with their own set of systematics and it is indeed unclear which would provide the more accurate calibration for actual science spectra.}%
,  calibrations do not provide a consistent wavelength solution, one can by no means expect to get for actual science data a wavelength calibration that is more accurate than the discrepancy between ThAr/FP and LFC wavelength solutions.

It is therefore clear that improvements to the wavelength calibration procedure are necessary to unleash the full potential of ESPRESSO. A more accurate wavelength solution is in particular needed for the fundamental constants project \citep[e.g.][]{Murphy2022}, but would certainly also be beneficial for other science cases like radial-velocity studies of exoplanets or the redshift drift experiment \citep{Sandage1965, Liske2008, Cristiani2023}.
The prime suspect for causing the discrepancies and inconsistencies highlighted in Figure~\ref{Fig:Comparison_ThArFP-LFC_Gauss} is the instrumental line-spread function (LSF). It is already well known that the ESPRESSO LSF is strongly non-Gaussian, different for the two slices and fibers, and varies significantly along individual spectral orders and across the detectors. In the following, we therefore develop a method to accurately determine the ESPRESSO LSF and to incorporate it in the wavelength calibration process.

\subsection{Measurement of the LSF}

\let\P\undefined
\let\S\undefined
\newcommand{\V }{ \boldsymbol{v} }
\newcommand{\Vk}{ v_k }
\newcommand{\DV}{ \Delta{}v }

\newcommand{\Y }{ \boldsymbol{Y}  }
\newcommand{\Yi}{             y_i }

\newcommand{\Ij}{             I_j }
\newcommand{\Cj}{             y^\textup{c}_j }
\newcommand{\Wj}{             {\sigma_j} }
\newcommand{\FINj}{           {\mathcal{F}_j} }

\newcommand{\Disp}{  \mathcal{D}  }
\newcommand{\D }{ \boldsymbol{F^\textup{obs}}  }
\newcommand{\Di}{             F^\textup{obs}_i }
\newcommand{\B }{ \boldsymbol{F^\textup{bg}}  }
\newcommand{\Bi}{             F^\textup{bg}_i }
\newcommand{\M }{ \boldsymbol{F^\textup{mod}}  }
\newcommand{\Mi}{             F^\textup{mod}_i }
\newcommand{\P }{ \boldsymbol{l} }
\newcommand{\Pk}{             l_k }

\newcommand{\A  }{ \boldsymbol{\mathsf{A}}   }
\newcommand{\Aik}{                  A_{ik}}

\newcommand{\f  }{   {f}  }

\newcommand{\S  }{  \boldsymbol{\mathsf{C}}^{-1} }
\newcommand{\C  }{  \boldsymbol{\mathsf{C}} }
\newcommand{\Cii}{  \sigma_i^2 }

\newcommand{\logL}{ \textup{log}\mathcal{L} }

\newcommand{\PrM}{ \boldsymbol{\mu}_{\P}^\textup{pr} }
\newcommand{\PrK}{ \boldsymbol{\mathsf{K}}_{\P}^\textup{pr}   }
\newcommand{\PrV}{ \boldsymbol{\kappa} }
\newcommand{\PrC}{ \boldsymbol{\mathsf{\Sigma}} }
\newcommand{\PrCnm}{ \mathsf{\Sigma}_{nm} }

\newcommand{\PoM}{ \boldsymbol{\mu}_{\P}^\textup{po} }
\newcommand{\PoK}{ \boldsymbol{\mathsf{K}}_{\P}^\textup{po}   }

The essential step in the whole process is of course to construct a model of the instrumental LSF. This is only possible if appropriate calibration exposures are available, composed of truly unresolved lines that reveal the instrumental LSF and allow to measure it accurately.
In the ideal case, this calibration source would provide infinitely narrow lines which are isolated and sufficiently separated to not cause blending, even in the extended wings of the LSF. At the same time, one needs many high signal-to-noise lines to accurately track the change of the LSF along all spectral orders.

\begin{figure*}
 \includegraphics[width=\linewidth]{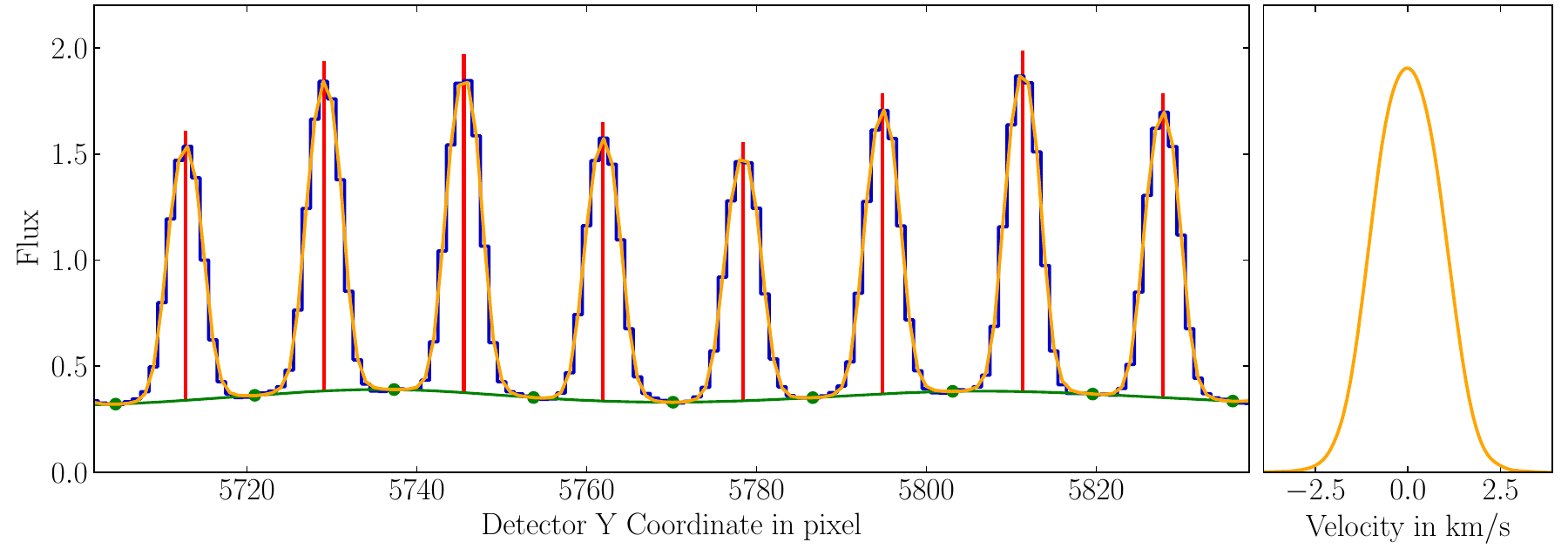}
 \caption{
  Visualization of the concept adopted to determine the LSF from LFC spectra (blue). In a first step, the level of the diffuse LFC background is estimated in between the lines and interpolated to create a smooth model of the background (green). Subsequently, the positions and intensities of individual lines are fitted and assumed to be intrinsically very narrow (vertical red lines). The LSF is then determined by finding a smooth function (right panel) which, when convolving the intrinsic LFC spectrum with said function, reproduces the observed data (blue).
 }
 \label{Fig:Visualization_Concept}
\end{figure*}

Spectra of the ThAr HCL in principle provide emission lines that should be fairly narrow. However, the lines in a thorium spectrum are unevenly bright, not regularly spaced, and often suffer from blending with other thorium or argon lines. This drastically limits the number of \textit{clean} lines that can be used and, given the demanding requirements for ESPRESSO, makes it impossible to derive a sufficiently high-quality LSF model from just the ThAr spectra.
The FP device, on the other hand, provides a plethora of stable, equally bright, and equally spaced lines. However, the ESPRESSO FP is designed with a relatively low finesse of $\mathcal{F} \approx 12$, leading to lines with an intrinsic full width at half maximum (FWHM) of $\approx0.8\;\kmps$ that appear in the extracted spectrum about 25\% wider than the instrumental LSF. This renders them unusable for an accurate LSF determination.
We therefore require spectra of the the LFC to measure the ESPRESSO LSF. This naturally limits this analysis and all subsequent steps to the spectral range covered by the LFC, i.e. $\gtrsim4300\,\AAA$.
The LFC provides a dense forest of truly unresolved lines with a width of $\approx120\,\textup{kHz}$ \citep{Probst2014}, corresponding to about $6\,\cmps$. Still, it is not the perfect source for LSF measurements and certain aspects have to be taken into account, for example, the substantial line-to-line flux variations (see e.g. Figure~\ref{Fig:Visualization_Concept}) and the line separation of $18\,\textup{GHz}$. The latter is at short wavelengths just sufficient to separate the lines, but does not provide any extra margins in between them.
Most critical within the context of LSF determination is, however, the diffuse background in the LFC spectra.
\hl{As described in detail e.g. by \citet{Ludwig2023}, diffuse background flux is a fundamental component of LFC spectra, depends strongly on frequency and the exact setup of the comb, and can actually contribute (over certain wavelength region) the majority of the photons in a LFC spectrum.}
\hl{For the comb installed at ESPRESSO}, the diffuse background increases towards shorter wavelengths and for $\lambda \lesssim 4500\,\AAA$ contains more photons than the actual LFC lines. It first has to be removed before the LSF shape can be determined.

Measuring the LSF and incorporating it in the wavelength calibration is therefore a fairly complex endeavor and requires determining the amplitude of the diffuse background, measuring the intensity and pixel position of every LFC line, inferring the instrumental line shape, and constructing the wavelength solution, i.e. the relation between pixel positions and wavelengths. Ideally, one would do all of this in one joint fit. This, however, would be far too complex and make the problem intractable. We therefore adopt a step-by-step approach, closely following the concept lined out by \citet{Hirano2020}, and solve each of these aspects independently.

\subsubsection{Background Determination}
\label{Sec:BackgroundDetermination}

First, we construct a model of the diffuse LFC background flux. For this, we find the minima in between individual lines and measure the flux in a small regions around them. These measurements of the diffuse background flux are then interpolated using cubic splines to get a smooth, continuous model, which is visualized in Figure~\ref{Fig:Visualization_Concept} (green dots and curve).

In principle, there can be a degeneracy between a true spectrally diffuse background and extended wings of the LSF that overlap and form a pseudo-background. Here, since we do not perform a joint fit but determine background and LSF profile separately, we can not distinguish between those two cases but require that neighboring lines are sufficiently far separated so that the contribution from the wings of the LSF in the middle between two lines is negligible and the observed flux at these positions actually representative for the true diffuse background. In the red part of the spectrum, where the LFC lines are, in terms of velocity and therefore relative to the FWHM of the instrument, further apart, this is no big problem and reliably determining the background in the described way is readily possible. For the blue end of the spectrum, this becomes more challenging, as can be seen in Figure~\ref{Fig:Visualization_Concept} which shows a part of the LFC spectrum around $4330\,\AAA$.
Still, we convinced ourselves that even there the adopted strategy for background estimation still provides sufficiently accurate values and at the moment does not pose a limitation for our analysis.

Nevertheless, we want to stress that this aspect of background estimation has to be kept in mind when designing a LFC for a given spectrograph. To be able to measure the diffuse background, the desired LFC line separation needs to be substantially larger than e.g. the formally optimal value for the line separation given in \citet{Murphy2007}.

\subsubsection{Initial Fit of Lines}

The next step in the process is to fit all LFC lines. The purpose of this is to derive preliminary pixel positions and line intensities which are required to infer the LSF. These information are visualized in Figure~\ref{Fig:Visualization_Concept} in red color.

The fitting process is straight forward. After subtracting the background model from the data, the procedure iterates over all LFC lines and fits each of them individually with a Gaussian line shape, delivering the centroid positions $\Cj$ and line intensities $\Ij$. This corresponds directly to the classical approach as it was also described in \citet{Schmidt2021}.
It was explored whether fitting a constant or even linear background in addition to the Gaussian line is necessary, but this was deemed to be not beneficial and the a~priori determined background model to be sufficient.
The fitted pixel positions can also be used to establish a preliminary wavelength solution. For the subsequent steps, it is necessary to convert from pixel positions to velocity space. Here, not a full wavelength solution but at least a dispersion relation, $\Disp$, giving the size of each pixel in $\kmps\:\textup{pix}^{-1}$ is needed. This does not need to be extremely precise since the regions of interest typically cover just a few pixels around the line centers. Thanks to the excellent stability of ESPRESSO, we can avoid computing a preliminary wavelength solution and instead, without jeopardizing accuracy, use for that purpose a static definition.

\subsubsection{Fit Procedure for the LSF}

The next, rather complicated step, deals with the determination of the LSF. For this, each echelle order is split into several blocks and the LSF measured individually for each of them. This provides maximum flexibility to model a variation of the LSF profile along each spectral order and from order-to-order or between fibers and slices. More, but smaller, blocks naturally allow for a more accurate but at the same time less precise determination of the LSF. As a compromise, we choose to split each order into 16 blocks, all 577~pixels long. These then typically contain between 16 and 40 LFC lines from which the LSF is inferred.

The concept for this is visualized in Figure~\ref{Fig:Visualization_Concept}. After constructing a model of the diffuse background (green) and fitting the intensity and pixel positions of all lines, we assume that the LFC lines are intrinsically extremely narrow (therefore indicated in Figure~\ref{Fig:Visualization_Concept} as vertical red lines). This knowledge about the intrinsic line profile is essential and we here rely on the measurement by \citet{Probst2014}, who demonstrated for a similar LFC system an intrinsic line width of less than $120\,\textup{kHz}$, about $6\,\cmps$.
The task is then to find a LSF model (right panel) that, when convolving the assumed intrinsic spectrum (composed of background and narrow lines) with the LSF, reproduces the observed data (blue).

We define the profile of the LSF in velocity space. This requires a dispersion relation to convert velocities to pixel coordinates, but allows to work with a more constant, less varying LSF. While the dispersion varies substantially along orders, from $365$ to $600\,\mps\:\textup{pix}^{-1}$, the resolving power $R = \lambda / \Delta{}\lambda$, and therefore the FWHM of the LSF expressed in $\kmps$, remains approximately constant along spectral orders. We therefore define the LSF on a velocity grid, $\V$, that extends over $\pm3.8\,\kmps$ in $\DV=10\,\mps$ wide bins. The LSF sampled on this grid is the model to be determined, composed of $761$ free parameters, which we index with the index $k$. Note that $k$ here runs from $-380$ to $+380$, so that $\Vk = k \cdot \DV$.
The setup of the velocity grid, i.e. extend and sampling, is to a certain degree arbitrary and reflects some compromise. Here, we adopt a an extremely fine sampling of the LSF, because we found some indications that better sampling slightly improves the results. In the interest of computational effort, one might want to pick a more economical bin size in the future. However, since only one dataset is processed within the context of this study, we here prefer to maximize accuracy.

\hl{To infer the LSF itself, we follow a fully non-parametric approach.
The blueprint for this has been lined out by \citet{Hirano2020}, who themselves follow the formulation of \citet{AsensioRamos2015}.}
The operation of relevance is a convolution between the assumed intrinsic spectrum and the to-be-determined LSF. After discretization, this can be expressed as a matrix-vector multiplication of the LSF profile, $\P = \{ \Pk \}$, with a matrix, $\A$, producing a model of the observed flux, $\M = \{ \Mi \}$,  as function of the pixel coordinates, $\Y=\{\Yi\}$, i.e.
\begin{equation}
 \Mi = \sum_k \Aik \, \Pk.
 \label{Eq:Model}
\end{equation}
The different columns of $\A$ all contain the assumed intrinsic spectrum, just shifted in velocity by multiples of $\DV$. To construct $\A$, the model for the intrinsic spectrum needs to be analytic (or sampled at extremely fine resolution and interpolatable).
We assume that it is composed of a set of narrow Gaussian LFC lines, $\f_\sigma(\Y)$.
Intensities $\Ij$ and centroid positions $\Cj$ for all lines contributing to the spectral chunk were fitted in the previous step and widths are assumed to be $\sqrt{8\ln(2)}\:\sigma = 6\,\cmps$, following \citet{Probst2014}. Therefore, the components of $\A$ can be computed as
\begin{equation}
\Aik \propto \sum_j \; \Ij \; f_\sigma\left( \Yi - \Cj - k\cdot\DV \cdot \Disp^{-1} \right) .
\label{Eq:Aik_FunctionEvaluation}
\end{equation}
Here, the dispersion relation, $\Disp$, is needed to convert from velocity space to pixel coordinates, as mentioned before.
Since the intrinsic width of the LFC lines (a few $\cmps$) is small compared to the sampling of the velocity grid, $\DV$, it is important to integrate the profile, $f_\sigma$, over the velocity bins, instead of just evaluating the function at the desired position. The exact formula for the components of $\A$ therefore becomes
\begin{equation}
\Aik = \sum_j \; \Ij \; \int_{\Yi-0.5}^{\Yi+0.5} f_\sigma\left( y - \Cj - k \cdot v \cdot \Disp^{-1} \right) dy .
\label{Eq:Aik_FunctionIntegration}
\end{equation}
Since we assume a Gaussian intrinsic line profile, this integration can be done analytically using the Gauss error function.
The diffuse background, on the other hand, is very smooth on the scales of the LSF. We therefore treat it separately from the convolution process, i.e. we directly subtract it from the observed spectrum and compare our background-free model $\M=\A\cdot\P$ to the background-subtracted flux $\D-\B$.
Following a classical least-squares approach, one can compute the likelihood, $\mathcal{L}$, as
\begin{equation}
\logL = -\frac{1}{2} \sum_i  \frac{ \left(\Di - \Bi - \sum_k \Aik \, \Pk \right)^2 }{\sigma_i^2 } .
\label{Eq:logL}
\end{equation}
Here, $\sigma_i$ denotes the uncertainty associated with flux measurement $\Di$. The corresponding variances can be summarized in one diagonal matrix $\C = \textup{diag}( \{ \Cii \} )$. Using some algebra, one can then demonstrate that the solution for $\P$ that maximizes $\logL$ (or minimizes $\chi^2$) is\hl{, analogous to Equation~4 in \citet{Hirano2020},} given by
\begin{equation}
\P^\textup{ml} = \left( \, \A^\top\S\,\A \, \right)^{-1} \: \A^\top \, \S \, \left( \, \D - \B \, \right) .
\label{Eq:MaximumLikelihood}
\end{equation}
Apart from the slightly complicated construction of $\A$, this is a straight-forward matrix operation. The matrix to be inverted in this process has the size of the LSF velocity grid squared, i.e. $761\times761$ for the adopted parameters.
However, directly computing the line profile in this way leads to an unacceptably large amount of noise. This is not unexpected.
Just considering the number of free model parameters (in this case 761) and observables in the input spectrum (here 577), it becomes clear that the problem is under-constrained. One could define the LSF on a less dense velocity grid and therefore reduce the number of model parameters, but this would not solve the problem. The main issue is the coarse sampling of the LFC lines in the observed spectra by just about four pixels across the FWHM (see Figure~\ref{Fig:Visualization_Concept}), while the model of the LSF needs to be specified on a much finer grid. There are numerous LFC lines which are fitted simultaneously and therefore provide sampling of the identical LSF with different pixel phases.
However, the pixel phases are defined by the dispersion relation of the spectrograph and the line spacing of the LFC, which can easily lead to redundancies. When (for a certain part of the LSF model) the pixel phases between different LFC lines are identical, they provide superfluous information and do not help to further constrain the LSF model.
Using multiple LFC calibration spectra would improve the photon noise but not be able to overcome the issue of limited sampling since the pixel phases are fixed and cannot be changed. It would be highly desirable to slightly shift the LFC lines between exposures and thereby obtain a vastly improved sampling of the LSF. Such tunability of the LFC is actively advertised by the manufacturer but, unfortunately, for the ESPRESSO LFC still not available in practice. Therefore, no useful additional data can be added to help obtaining a better LSF model.

Fortunately, there is prior knowledge that can be added to the inference problem and will massively help  to better constrain the model.
One might certainly expect the LSF to have an overall rather complicated shape, but it should still be smooth on small scales.
It therefore makes sense to add this prior information in the calculation of the LSF. Within a Bayesian framework, this is readily possible.
The likelihood $\mathcal{L}$ given in Equation~\ref{Eq:logL} represents the probability of the observations $\D$ given a certain LSF model, i.e.
\begin{equation}
\mathcal{L} = p(\,\D\,|\,\P\,) .
\end{equation}
The fundamental principle of Bayesian inference is that the posterior probability for the quantity of interest, in our case $p(\,\P\,|\,\D\,)$, can be expressed as function of the likelihood and some prior. For the problem at hand, this can be written as
\begin{equation}
p(\,\P\,|\,\D\,) \propto p(\,\D\,|\,\P\,) \: p(\,\P\,).
\label{Eq:Bayes}
\end{equation}
Here, the prior $p(\,\P\,)$ allows to include extra knowledge about the LSF.
The maximum-likelihood approach described in Equation~\ref{Eq:MaximumLikelihood} corresponds to the special case in which no prior knowledge is taken into account and the individual values of $\P$ are a~priori fully unconstrained and uncorrelated.
This, however, does not properly reflect the situation we have in practice.
In fact, he have quite substantial prior knowledge about the instrumental line shape.
We know that the LSF should resemble something like a line, i.e be localized with high flux in the center of our velocity grid and more or less asymptotically approach zero in the wings and far away from the center. We even have some estimate about its width and shape: The classically assumed Gaussian line shape is certainly not accurate, but also not completely wrong and can serve as a first estimate. In addition, the LSF should by definition be normalized and never be negative. Most crucially, however, we want to enforce it to be smooth, i.e. with little to no fluctuations on the smallest scales.

A very clever and convenient way for adopting priors is the approach lined out in \citet{Hirano2020}, which turns the full LSF inference problem into a Gaussian process regression.
For this, the prior distribution has to be of the form
\begin{equation}
 p(\,\P\,) = \mathcal{N}[\,\P\,|\,\PrM,\PrK\,] ,
 \label{Eq:Prior}
\end{equation}
where $\mathcal{N}$ represents a multivariate Gaussian distribution, characterized by a prior mean, $\PrM$, and a covariance matrix $\PrK$.
Therefore, each prior realization of $\P$ is a draw from a multivariate Gaussian distribution with given properties, described by the given mean and covariance.
Since $\P$ is only defined and sampled on a given discrete velocity grid, the Gaussian distribution $\mathcal{N}$ has only finite dimensionality, which makes the problem computable in practice.

An essential aspect in this process is of course the choice of $\PrM$ and $\PrK$, which encapsulates all prior knowledge and desired behavior of the LSF model. \citet{Hirano2020} chose a vanishing prior mean and a classical squared-exponential kernel with some correlation length to define the covariance matrix. We try to improve upon this with a slightly more complex and more informative scheme, which will be described in detail below.
Nevertheless, it is worth to check which of the properties known about the LSF can at least in principle be specified by this Gaussian process prior. Here, one notices that the constraint of non-negativity is difficult to enforce. For a Gaussian process, the prior (but also the posterior) probability for every given point in the model, i.e. in our case corresponding to $p(\,\Pk\,)$ or $p(\,\Pk\,|\,\D\,)$, is always a Gaussian distributions. Far away from the core of the line, the values of the LSF model are expected to be quite low and ideally approach zero. Also enforcing them to be non-negative will unavoidably lead to a strongly asymmetric and therefore non-Gaussian probability distribution. Such a behavior can not be realized within a Gaussian process framework. The usual tricks to get around this limitation, e.g. taking the logarithm of the flux or working on its absolute value, are not appropriate here either.
Still, the benefits of a Gaussian process approach prevail this limitation.

The biggest advantage of the adopted approach is that the posterior probability can be obtained in a rather simple, fast, and efficient way.
The adopted likelihood (Equation~\ref{Eq:logL}) is of the classical least-squares form and therefore a multivariate Gaussian. The prior (Equation~\ref{Eq:Prior}) is also a multivariate Gaussian. This allows to solve analytically for the posterior distribution, which turn out to be as well a multivariate Gaussian and is therefore fully described by its mean and covariance.
Combining equations~\ref{Eq:logL}, \ref{Eq:Bayes}, and \ref{Eq:Prior}, one obtains for the posterior
\begin{equation}
\begin{split}
p(\,\P\,|\,\D\,)    &= \mathcal{N}[\,\D-\B\,|\,\A\cdot\P,\C\,] \cdot \mathcal{N}[\,\P\,|\,\PrM,\PrK\,] \\
                    &= \mathcal{N}[\,\P\,|\,\PoM,\PoK\,] ,
\end{split}
\end{equation}
which reduces the problem to determining the posterior mean, $\PoM$, and covariance, $\PoK$.
Again, with some effort, one can demonstrate that the posterior covariance is given by
\begin{equation}
 \PoK = \left( \: \A^\top \, \S \, \A + \left(\,\PrK\,\right)^{-1} \: \right)^{-1}
 \label{Eq:PosteriorCovariance}
\end{equation}
and the posterior mean by
\begin{equation}
 \PoM = \PoK \, \left( \: \left(\,\PrK\,\right)^{-1} \PrM \, + \, \A^\top \, \S \, \left( \, \D - \B \, \right) \: \right) ,
 \label{Eq:PosteriorMean}
\end{equation}
\hl{which is as well fully equivalent to Equations~11 and 12 in \citet{Hirano2020}.} Therefore, one can obtain a full description of the posterior distribution of $\P$ by just some matrix operations of which the heaviest are the two inversions in Equation~\ref{Eq:PosteriorCovariance} related to the prior and posterior covariance. Their dimensionality, as stated before, corresponds to the size of the LSF velocity grid squared, i.e. $761\times761$. The inversion of $\C$ is trivial since it is diagonal.
Nevertheless, significant computations are also needed to construct $\A$.
Equations~\ref{Eq:PosteriorCovariance} and \ref{Eq:PosteriorMean} resemble Equation~\ref{Eq:MaximumLikelihood}, which is only logical, since the latter corresponds to the special case of vanishing priors.

Gaussian processes are often presented with vanishing prior mean. However, we feel that this would not properly reflect our a~priori knowledge.
We therefore adopt as prior mean a Gaussian LSF with a width corresponding to the nominal resolution of ESPRESSO. Although not fully accurate, this should be closer to the truth than assuming a~priori that the LSF is zero everywhere. Including all necessary normalization terms, the prior mean can be expressed as
\begin{equation}
 \PrM(\V) = \sqrt{\frac{8\log(2)}{2\pi\,{L_\mu}^2}} \; \DV \cdot \textup{exp} \left[ {-\frac{1}{2}\frac{ 8\log(2) \;   \boldsymbol{v}^2 }{ {L_\mu}^2} } \right] .
 \label{Eq:PriorMeanExplicit}
\end{equation}
Here, $L_\mu$ specifies the FWHM of the profile, for which we adopt a value of $2.3\,\kmps$, corresponding to a nominal resolving power of $R=130\,000$.

\begin{figure}
 \includegraphics[width=\linewidth,page=6]{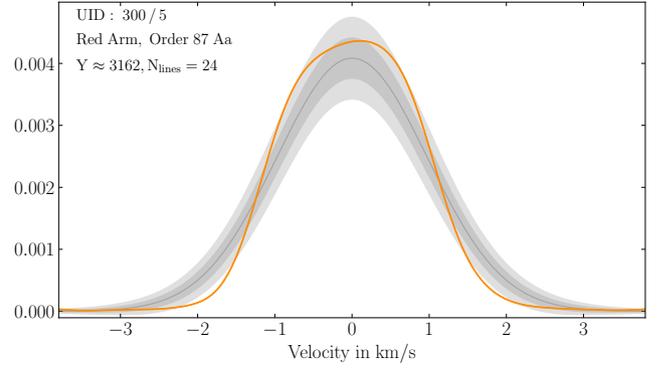}
 \caption{
  Visualization of the fitted LSF model for a single part of one spectral order (Order~87, Fiber~A, Slice~a, Block~5 of 16). The determined LSF profile is shown in orange. As comparison and to highlight the non-Gaussianity and skewness of the determined line profile, the prior mean and variance are shown in gray. Grey shaded areas indicate the 1-$\sigma$ and 2-$\sigma$ region. Note, that the dependence on the prior mean is eliminated during the fit procedure by iteration over Equation~\ref{Eq:PosteriorMean}.
 }
 \label{Fig:LSF1+2}
\end{figure}

\begin{figure}
 \centering
 \includegraphics[width=.8\linewidth]{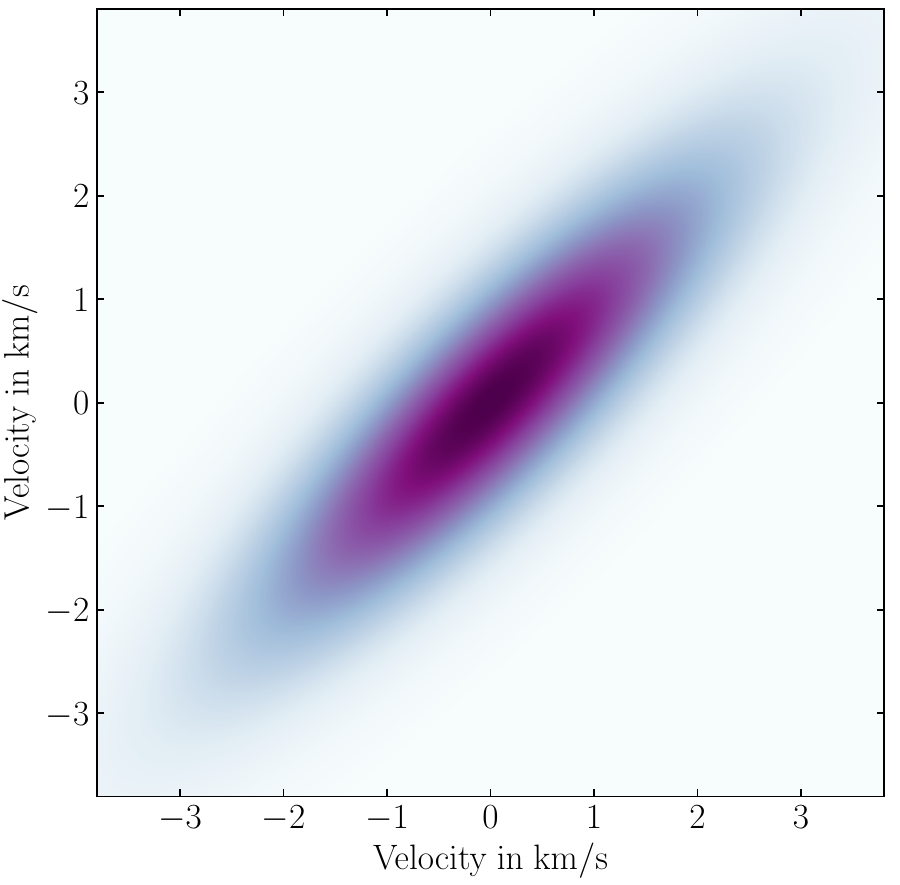}
 \caption{
  Two-dimensional representation of the adopted prior covariance matrix. For better visualization, the squareroot of $\PrK$ is shown.
 }
 \label{Fig:CovarianceMatrix}
\end{figure}

Also for the prior covariance, we try to include as much physical knowledge as possible.
We therefore adopt a form in which the covariance matrix can be decomposed as
\begin{equation}
 \PrK = \textup{diag}(\PrV)^\top \; \PrC \; \textup{diag}(\PrV)
 \label{Eq:PriorCovarianceExplicit}
\end{equation}
into a correlation matrix, $\PrC$, and the prior variance, $\PrV^2$.
For the correlation matrix, we assume a standard squared-exponential kernel of the form
\begin{equation}
 \PrCnm = \textup{exp} \left[ {-\frac{1}{2}\frac{ 8\log(2) \; ( \, v_n \, - \, v_m )^2 \, }{{L_\Sigma}^2} } \right]
 \label{Eq:PriorCorrelationExplicit}
\end{equation}
in which $L_\Sigma$ specifies the correlation length, again expressed as the FWHM of the kernel.
Assuming $\PrV$ to be constant would turn this into the default stationary squared-exponential kernel.
However, we expect the LSF to approach zero far away from the core of the line and therefore do not want to grant the model much freedom in this region. Close to the center, on the other hand, the flux is much higher and the model needs to be flexible to capture the structure of the LSF. This behavior naturally demands a non-stationary kernel. Therefore, we provide a prior variance that allows large variations around the prior mean in the core of the LSF and only small ones in the wings.
The exact form we adopt is
\begin{equation}
 \PrV(\V) = \sqrt{\frac{8\log(2)}{2\pi\,{L_\mu}^2}} \; \DV \cdot \left( \, \sigma_0 + \sigma_f \cdot \textup{exp} \left[ {-\frac{1}{2}\frac{ 8\log(2) \;   \V^2 }{ {L_\kappa}^2} } \right] \, \right) .
 \label{Eq:PriorVarianceExplicit}
\end{equation}
The width of the prior therefore follows a Gaussian profile, characterized by $L_\kappa$ and $\sigma_f$. To not over-constrain the model in the wings of the line, we add to this a constant term, $\sigma_0$.
The normalization factor up front is just for convenience and allows to specify the amount of prior uncertainty, i.e. $\sigma_0$ and $\sigma_f$, in a handy format relative to the peak of the prior mean.
The resulting prior is visualized in Figure~\ref{Fig:LSF1+2}. As gray shaded areas, it displays the $1\,\sigma$ and  $2\,\sigma$ region allowed by the prior variance around the prior mean. It also shows the posterior mean obtained in the described way. One can notice how the inferred line profile deviates from a Gaussian and also exhibits some asymmetry. A two-dimensional representation of the full covariance matrix, $\PrK$, is given in Figure~\ref{Fig:CovarianceMatrix}.

As visualized in Figure~\ref{Fig:LSF1+2}, the chosen prior is quite narrow. This ensures a robust and stable determination of the LSF, even when the data quality varies. However, one might be concerned that this quite stringent constraint pushes the inferred LSF too much towards the initially assumed Gaussian profile. To avoid any bias by the prior mean, we execute the inference process multiple times, using the posterior mean of the previous iteration as the prior mean of the subsequent one, while keeping the covariance fixed. Fortunately, this is quite simple, one only has to iterate over Equation~\ref{Eq:PosteriorMean}. This does not require re-computation of neither the large matrix $\A$ nor new matrix inversions and can therefore be executed quickly. In this way, the posterior converges towards the final result which is virtually independent of the initial prior mean, even when the prior covariance constrains the process relatively tightly%
\footnote{One might argue why any informative prior mean is necessary in the first place if during the iteration process any dependence of the final posterior on initial prior mean is eliminated. An uninformative prior might indeed perform equally well, but a direct comparison with an uninformative prior was not conducted. Instead, we adopted at some point of the study the described approach using an informative prior mean, implemented the iteration process, and validated this concept. We still expect that this provides the Gaussian process some additional guidance, but we do not claim that it is necessary.}.
At the same time, the iterative process allows to quantify the influence of the prior mean. By examining the result of each individual iteration, one can see how the first posterior is still biased towards the initial Gaussian prior mean, but subsequent iterations converge very quickly towards the final result with only extremely small and insignificant changes from iteration to iteration.
This demonstrates that our procedure does not bias the inferred LSF.

\begin{table}
	\centering
	\caption{Summary of the hyperparameters and their value adopted  for the inference of the ESPRESSO LSF.}
	\label{Tab:Hyperparameters}

\begin{tabular}{cc} \hline
 Hyperparameter                           & Value                 \\ \hline
 $\V_\textup{min} / \V_\textup{max}$      & $\pm3.8\,\kmps$       \\
 $\DV$                                    & $0.010\,\kmps$        \\
 $N_\textup{blocks}$                      & 16                    \\
 $L_\mu$                                  & $2.3\,\kmps$          \\
 $L_\Sigma$                               & $1.5\,\kmps$          \\
 $L_\kappa$                               & $3.5\,\kmps$          \\
 $\sigma_0$                               & $0.002$               \\
 $\sigma_f$                               & $0.080$               \\ \hline

 \end{tabular}
\end{table}

The adopted scheme for determining the LSF model has in addition to the definition of the velocity grid six hyperparameters. These are listed in Table~\ref{Tab:Hyperparameters}. We do not attempt to formally optimize these. As will be shown below, the data contains systematics beyond the formal uncertainties. In such a case, a purely formal optimization is not guaranteed to deliver the desired results. Therefore, we pick by hand a set of hyperparameters that return satisfactory results. Also, we adopt the same parameters for every spectral order, independent of the wavelength. Nevertheless, we went through a quite extensive optimization process in which we manually modified the value of each hyperparameter within a certain range. It turns out that the quality of the obtained wavelength calibration does not significantly depend on the choice of the hyperparameters. They therefore seem to be close to optimal and we are quite certain that the systematics and discrepancies that are still present in the wavelength solutions (see sections below for a detailed discussion) can not simply be solved by a further optimization of the hyperparameters. The exact value of the hyperparameters is therefore at the moment not the limiting factor.

\begin{figure*}
 \includegraphics[width=\linewidth,page=7]{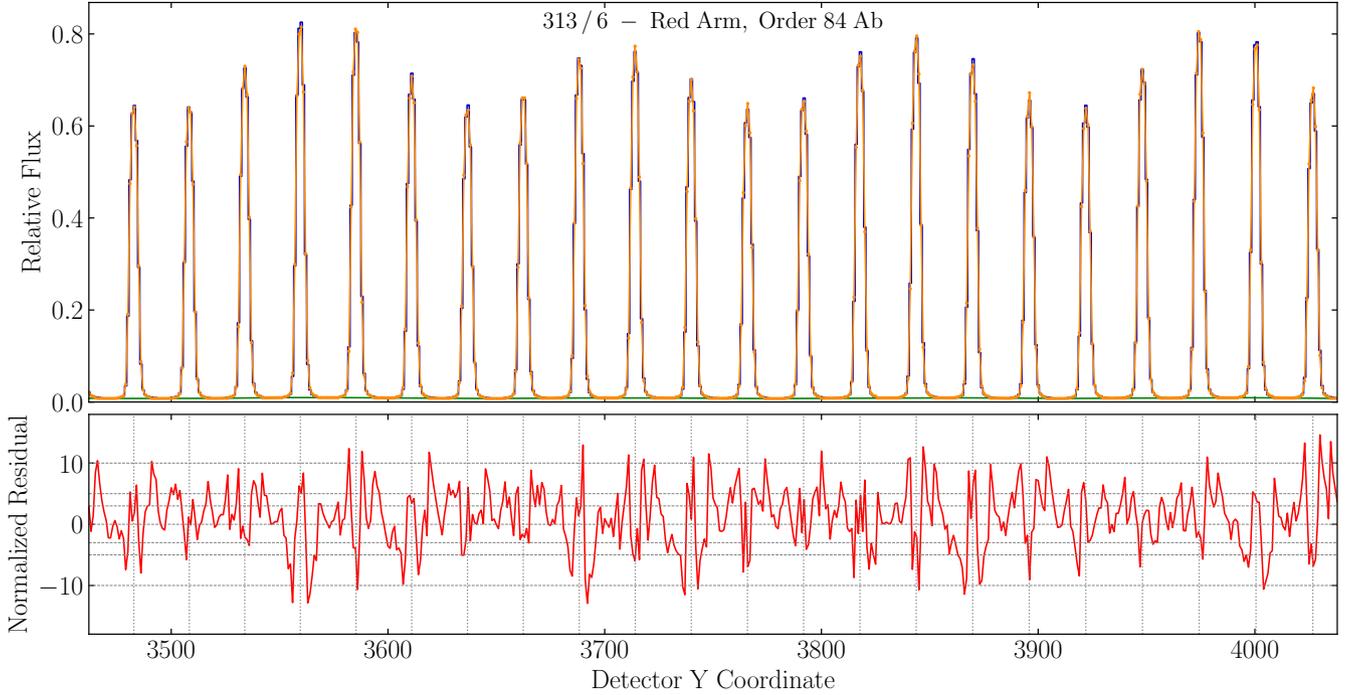}
 \caption{ Comparison between model and data.
  The top panel shows, for a small section of one spectral order, the observed flux in a LFC spectrum (blue), as well as the obtained model of the diffuse LFC background (green). In addition, it displays the flux predicted by the LSF model (orange).
  The bottom panel shows the error-normalized residuals between data and model. Horizontal lines indicate deviations by $3$, $5$, and $10\,\sigma$, while vertical lines mark the fitted initial positions, $\Cj$, of the individual LFC lines.
 }
 \label{Fig:Flux_ModelData}
\end{figure*}

One very important check is of course to see how well the inferred model fits the actual data.
The mean predicted model flux can easily be obtained by feeding the posterior mean (Equation~\ref{Eq:PosteriorMean}) into Equation~\ref{Eq:Model}. Similarly, information about the model uncertainty can be computed using the model variance, i.e. the diagonal of the posterior covariance in Equation~\ref{Eq:PosteriorCovariance}.
A comparison of this model flux to the observed flux is shown in Figure~\ref{Fig:Flux_ModelData}. The top panel shows, for a small section of one spectral order, the observed flux in a LFC spectrum, as well as the model of the diffuse LFC background. In addition, it displays the predicted model flux.
The LSF-based model obviously fits the data quite well and definitely much better than a set of purely Gaussian lines.
Nevertheless, one can notice some discrepancies. These are at a low level, but still statistically significant, because the S/N in this LFC calibration exposure is naturally very high.
The bottom panel therefore shows the error-normalized residuals between data and model. One can identify characteristic residuals: For the line at pixel position 3560, one can see a positive deviation of the data w.r.t. the model at the peak of the line, which turns into a quite significant negative deviation in the flanks of the line. The same pattern can be seen at $Y=3610$. The line in between, however, shows a pattern that is nearly identical, but inverted. In fact, similar residuals can be seen for many LFC lines (in this and other orders), either inverted or not. This demonstrates that the discrepancy between data and model is not simply due to a lack of flexibility in the model. Instead, a much more fundamental assumption of the modeling approach seems to not be satisfied: Apparently, neighboring lines do not exhibit the identical line profile and modeling them with an identical LSF leads to the shown residuals.
At the moment, not much can be done about this. With the existing calibration setup, the ESPRESSO LSF can only be measured at the given positions of the LFC lines. An effect that obviously changes from one LFC line to the next, however, needs to be characterized on much smaller scales, i.e. less than the separation of two LFC lines. This would only be possible with a tunable LFC. Since this is not available, there is little room for further investigations. We just note that the so called \textit{beat pattern noise} described in \citet{Schmidt2021} also had a period of about two LFC lines and--despite all differences--has some similarities with what we see here. It might be possible that there is a common origin to both effects.

One also has to note that the initial pixel positions of the LFC lines, $\Cj$, that went into the construction of $\A$, were obtained assuming a Gaussian line profile. Given the strongly non-Gaussian line profile of ESPRESSO (see e.g. Figure~\ref{Fig:LSF1+2}), one has to expect imperfections in these fits. However, re-fitting the line positions with the more correct non-parametric LSF model and performing the full LSF determination again with these improved line positions only leads to slightly cleaner residuals and overall a slightly better fit of the model but not to any substantial changes in the results.

\subsection{Determined LSF Profiles}

\begin{figure*}
 \includegraphics[width=\linewidth]{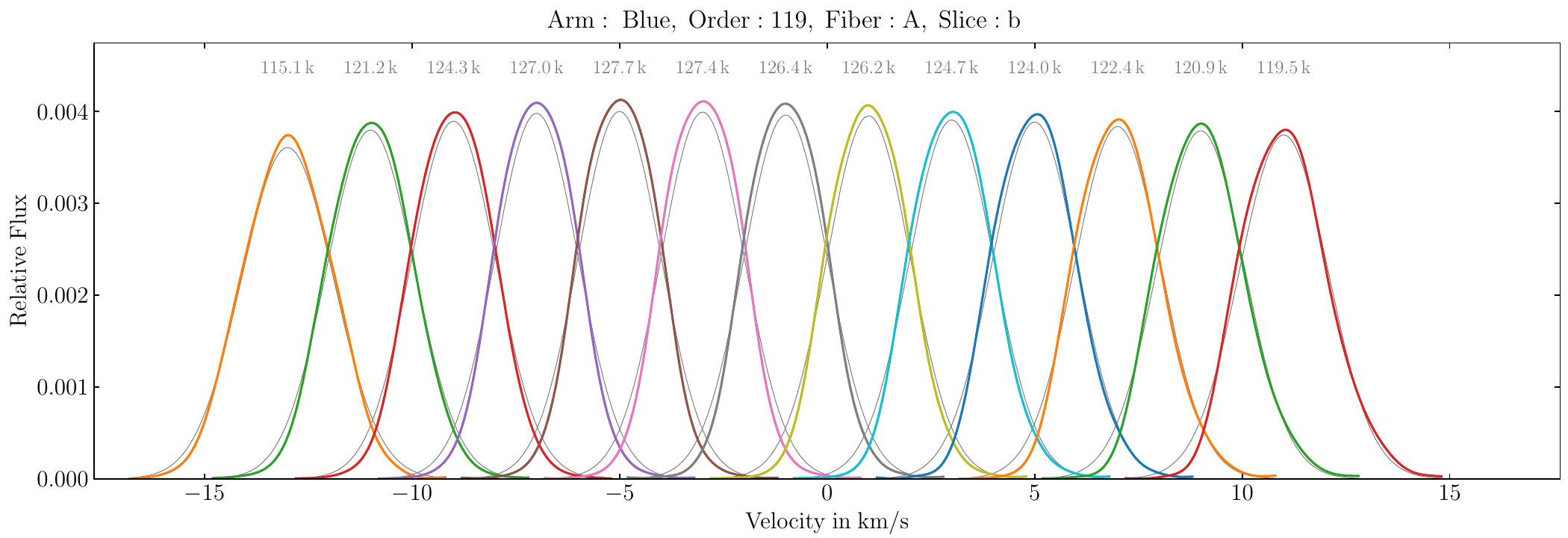}\\
 \includegraphics[width=\linewidth]{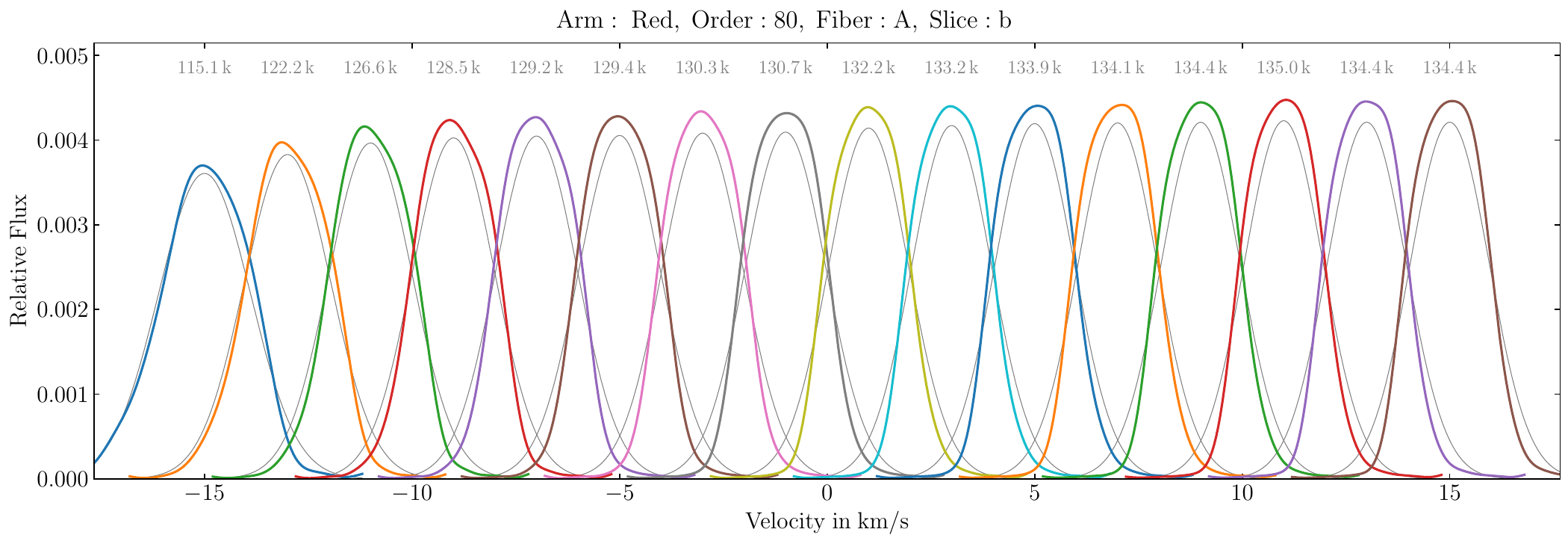}
 \caption{
  Gallery of LSF models obtained for two selected orders (Order 119, Fiber~A, Slice~b, central wavelength $\textup{5142\,\AAA}$, top panel, and Order~80, Fiber~A, Slice~b, central wavelength $\textup{7648\,\AAA}$,bottom panel).
  Individual LSF models are displaced horizontally for visualization purpose. For comparison, Gaussian profiles with the same FWHM as the empirical profiles are shown in gray. The FWHM, expressed as resolving power, are stated at the top. In the top panel, only 13 profiles are shown since the outermost regions of the blue detector receive no light.
 }
 \label{Fig:LSF_Profiles_169+329}
\end{figure*}

After inferring the ESPRESSO LSF profiles using the approach described above, it is of course interesting to examine the results and verify that the determined models agree with the expectations. This is quite a wealth of information. ESPRESSO contains 78 spectral orders%
\footnote{The LFC spectra only cover about 64 orders with sufficient flux.}%
, each imaged in two fibers and two spectral slices, and we determine the LSF in up to 16 independent blocks per order. This results in about 4000 individual LSF profiles. An ensemble of them from two selected spectral orders is shown in Figure~\ref{Fig:LSF_Profiles_169+329}.

As outlined before, the adopted Gaussian process scheme does not allow to enforce a strictly positive LSF.
Also, in Section~\ref{Sec:BackgroundDetermination}, we described the difficulty of disentangling the diffuse LFC background from overlapping line wings and our approach to not fit both aspects simultaneously but in separate steps. Therefore, if the background estimate is too high, the LSF model will try to compensate and become negative far away from the line core, or positive if the background estimate is too low.
Closely examining the determined LSF profiles reveals that this is not the case. The behavior in the outskirts of the LSF is as expected. It approaches zero and shows some scatter around it. In consequence, some negative values appear from time to time, but there is no indication that the wings of the LSF are systematically biased to compensate for an incorrect background.
We also manually introduced a re-scaling of the background level to deliberately provoke such a behavior. Although only tested for a few selected positions within the full spectral range, we got the impression that the determined background levels are in general not systematically biased, neither high nor low, and also rather precise. Re-scaling the background level by more than two per-cent up or down lead to substantially worse LSF models with quite visible deviations from zero in the outermost parts of the profile. We therefore conclude that the adopted scheme of first determining the background in between the lines and then constructing the line profile is certainly not optimal but good enough for our purpose at the moment.

Another property of the LSF models that is not actively enforced is their normalization.
The LSF model will naturally adjust itself to produce the average flux level that makes the model flux, $\M$, most consistent with the data, $\D - \B$. If this flux is the same as contained in the initial intensity estimates of the LFC lines, $\Ij$, the integral of the LSF will be unity.
We check for this and find only very minute (less than 1\%) deviations from unity. Given that the profiles are only defined on a velocity grid with some given extent, i.e. with finite support, some deviations from unity are expected.
We therefore do not introduce any further steps to normalize the LSF profiles.

The most interesting aspect of the LSF profiles is of course their shape and how they deviate from a simple Gaussian. Some examples are shown in Figure~\ref{Fig:LSF_Profiles_169+329}. One can notice the great variety of shapes and also how significantly the profiles change along orders. To highlight the non-Gaussianities, we plot (in gray color) for each empirically determined LSF also a Gaussian profile with the same FWHM.
In particular for the right part of spectral Order~80 (bottom panel), one can see that the LSF profiles are quite \textit{boxy}, i.e. they are relatively flat at the top and then fall off much steeper than a comparable Gaussian. This is not completely unexpected, given that the two-dimensional instrumental profile results from the convolution of the flat illumination pattern at the fiber exit facet with some blurring related to optical aberrations. Here, the intrinsic optical quality of ESPRESSO is so good that the contribution by the fiber core becomes quite noticeable in the LSF profile. For the left part of Order~80, but also for Order~119 (top panel), the blurring by the optics seems to be more dominant, resulting in an overall broader and more Gaussian line profile.
Another effect is visible in the right half of Order~119. Here, the asymmetries in the line profile become quite apparent. While one flank of the LSF drops steeper than a comparable Gaussian, the other side is substantially more extended. This is also visible for the left-most profiles in Order~80.
These deviations from a Gaussian line profile are of course not huge, but they can have a very significant impact on the science results, given that we want to measure extremely small effects. Just as a reference, an already relevant shift of just $2\,\mps$ would correspond to less than $\sfrac{1}{1000}$ of the FWHM of the line profile. At this point, even barely visible deviations of the LSF from a Gaussian shape can not be ignored.

\begin{figure*}
 \includegraphics[width=\linewidth]{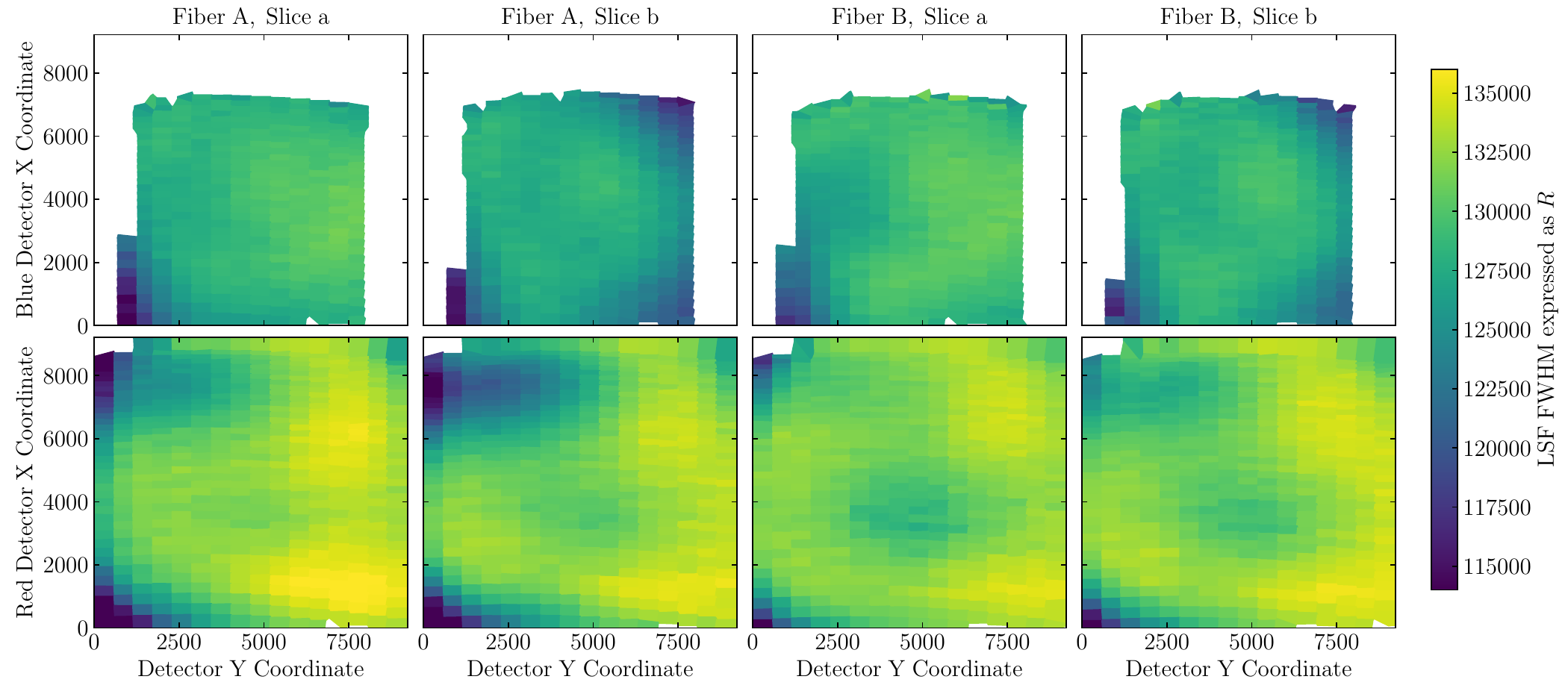}
 \caption{
  Map of the LSF FWHM as function of position on the ESPRESSO detectors. Top and bottom row of panels correspond to blue and red detector, respectively. The different columns depict results obtained from the different fibers and slices. In this format, the individual spectral orders run from left to right, with the bluest orders at the top and the reddest ones at the bottom. The empty space at the left and right edge of the blue detector are related to the optical design of ESPRESSO. These parts of the blue detector are unused and receive no light. The blank space at the top is due to the limited spectral coverage of the LFC, which provides insufficient flux for $\lambda\lesssim4300\,\AAA$.
 }
 \label{Fig:Map_FWHM}
\end{figure*}

Instead of inspecting individual LSF profiles, one can also compute different kinds of summary statistics and then explore how these change e.g. as function of wavelength.
The probably most straight-forward quantity to compute is the FWHM derived in a non-parametric fashion directly from the individual LSF models. Expressed as resolving power $R=\frac{\lambda}{\Delta{}\lambda}$, this property was already given in Figure~\ref{Fig:LSF_Profiles_169+329} for the profiles shown there. In Figure~\ref{Fig:Map_FWHM}, we visualize the FWHM derived from all fitted LSF models and display it as function of detector position. The top row of panels corresponds to the blue arm of ESPRESSO and the bottom one to the red arm. The four columns are related to the two fibers and two slices of ESPRESSO. The location on the detectors for these is of course nearly identical, but due to some non-common optical elements and slightly different light paths, all fibers and slices exhibit individual and often substantially different LSF profiles.
Most obvious in Figure~\ref{Fig:Map_FWHM} is that FWHM and corresponding resolving power are not constant across the detectors, but vary substantially by over 20\% in a non-trivial way. For instance, on the red detector, the region of highest resolving power approximately follows an inverted 'S'-shape with some regions of lower resolution on the top-left, approximately around $(X|Y)=(7000|2500)$, and bottom-right, near $(X|Y)=(3000|7000)$. However, the exact location and strength of this pattern is different among the fibers and slices. For example, the region of lower resolution in the top-left is most pronounced for Fiber~A, Slice~b and only very moderately present for Fiber~B, Slice~a.

\begin{figure*}
 \includegraphics[width=.49\linewidth]{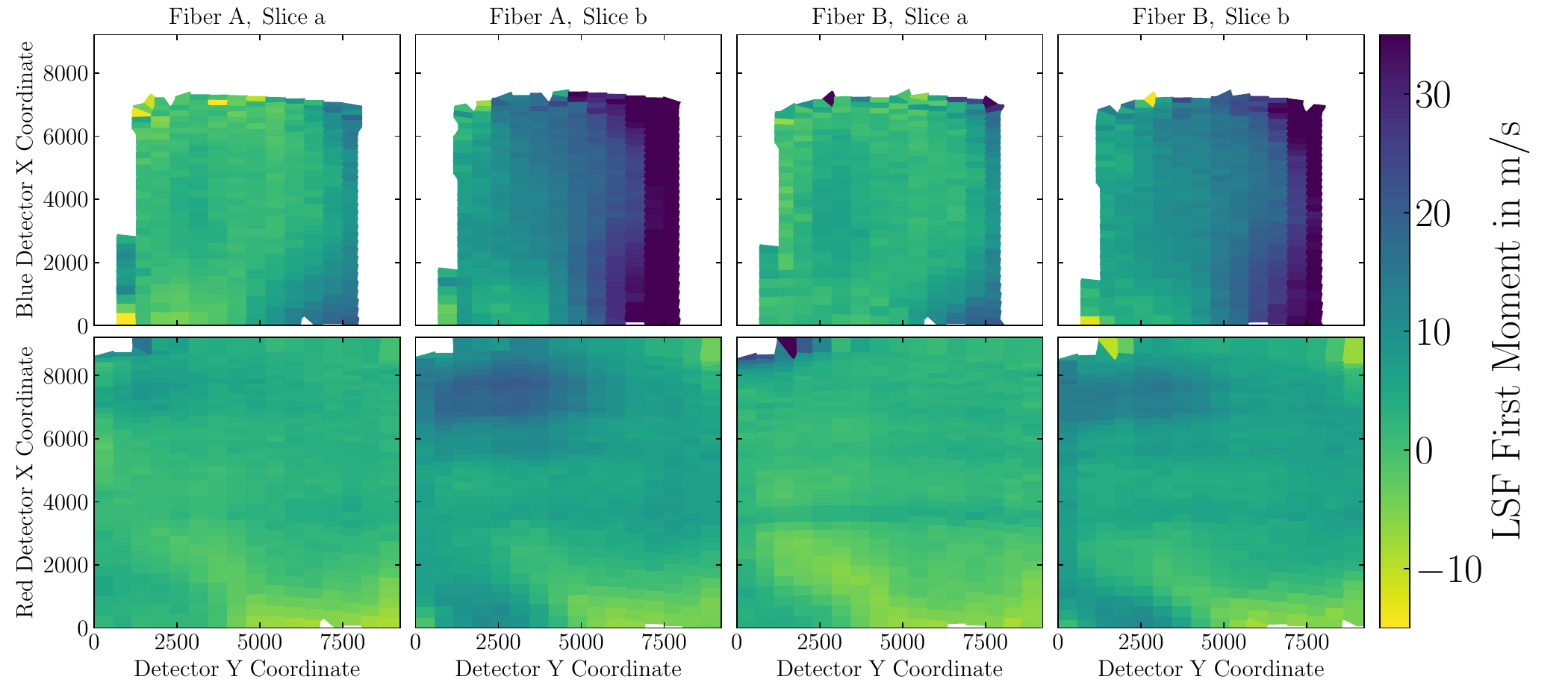}
 \includegraphics[width=.49\linewidth]{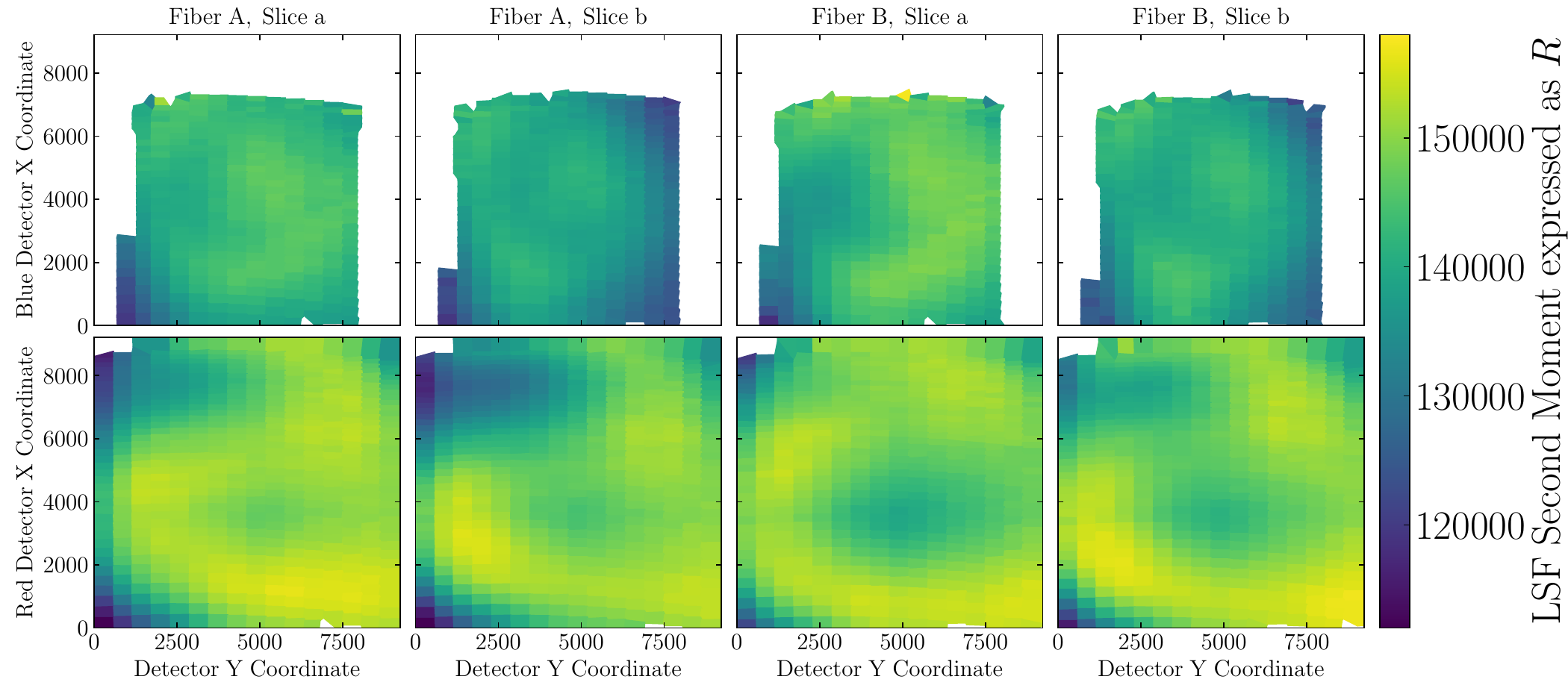}
 \includegraphics[width=.49\linewidth]{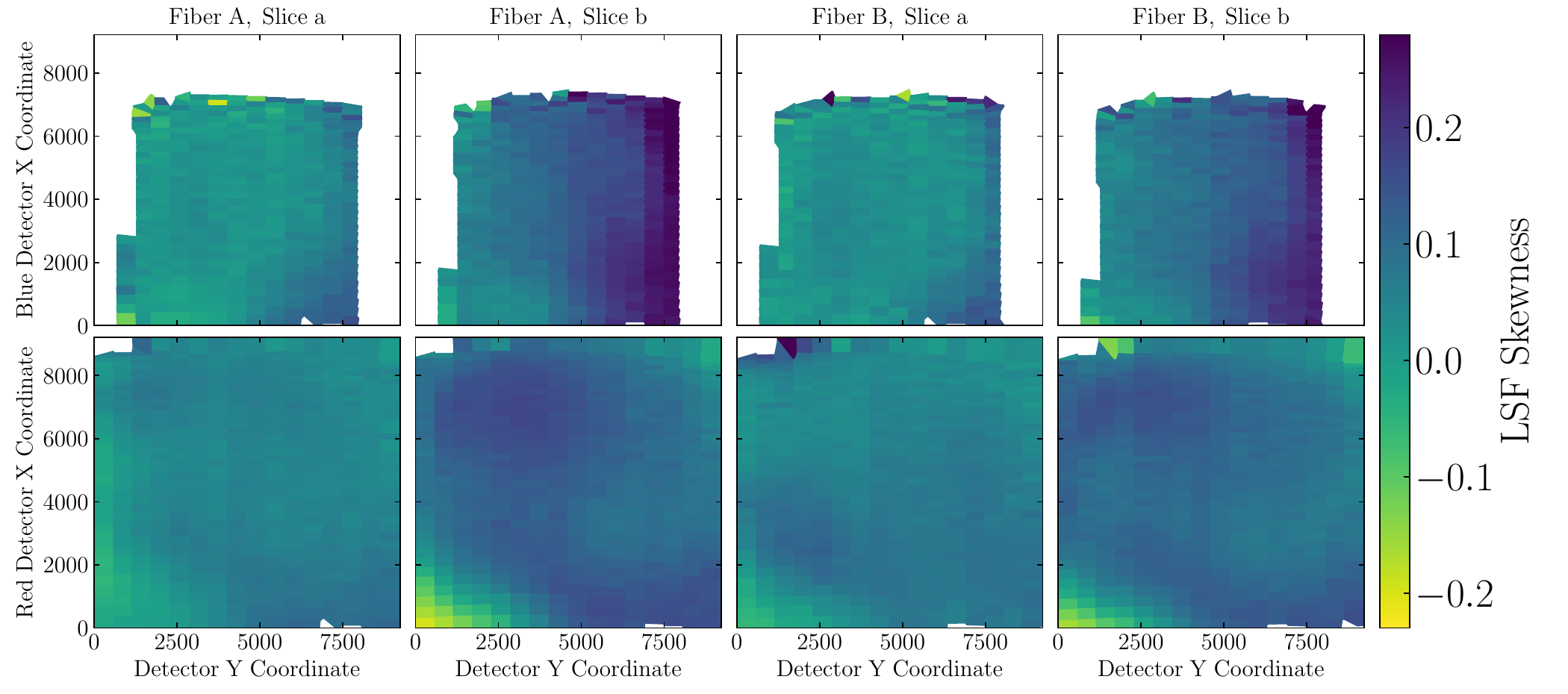}
 \includegraphics[width=.49\linewidth]{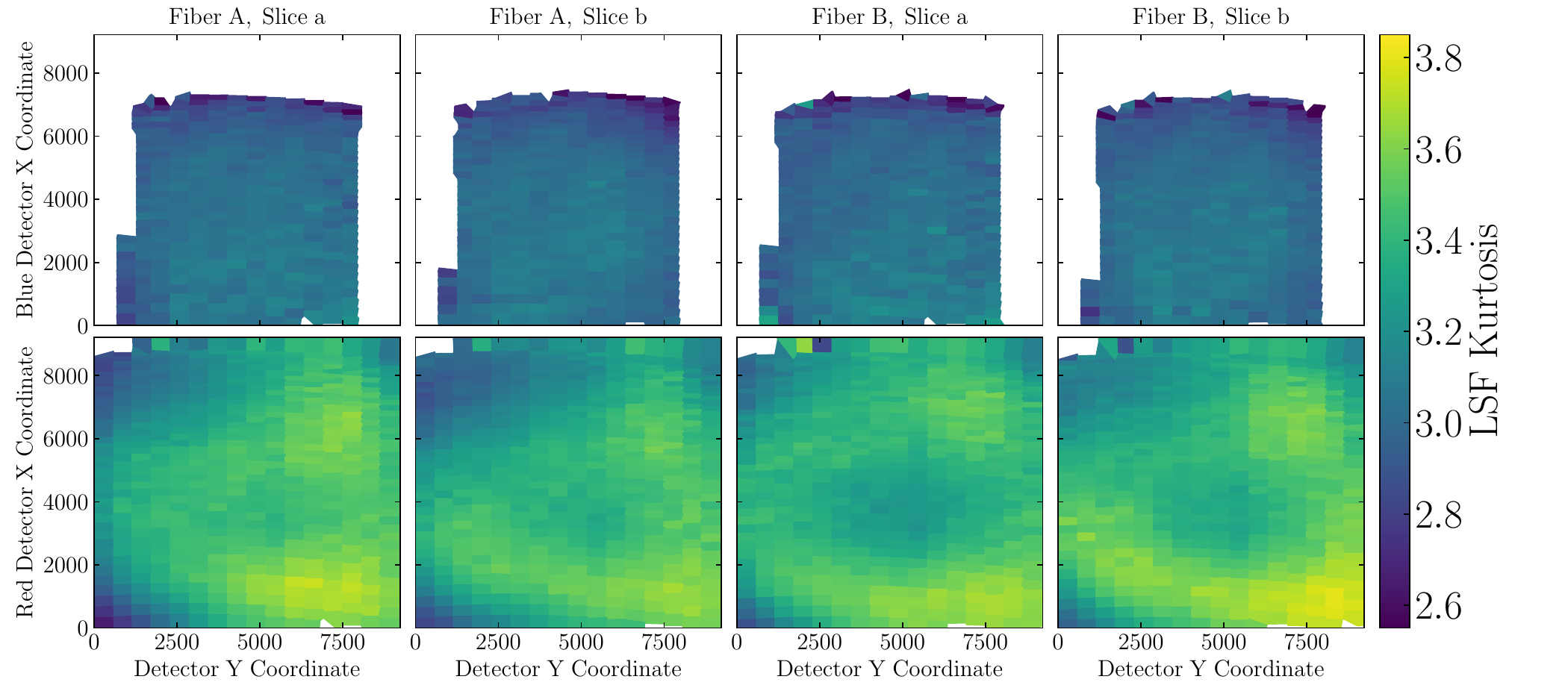}
 \caption{
  Similar to Figure~\ref{Fig:Map_FWHM}, but showing properties derived from the different moments of the LSF profile. The top-left panel shows the first moment, the top-right one the second moment, expressed as resolving power, the bottom-left panel the skewness and the bottom-right one the kurtosis of the LSF.
 }
 \label{Fig:Map_Moments}
\end{figure*}

In Figure~\ref{Fig:Map_Moments}, we display further properties of the LSF derived from different moments.
The top-right panel displays the resolving power computed from the second moment of the profiles. By construction, this is very similar to the FWHM measurement shown in Figure~\ref{Fig:Map_FWHM}, but there are subtle differences and the resolving power computed from the second moment is in general higher. This discrepancy comes from the fact that the actual ESPRESSO LSF is not a Gaussian. Therefore, the true FWHM (Figure~\ref{Fig:Map_FWHM}) is different from the FWHM one obtains when modeling the lines with a Gaussian profile (Figure~\ref{Fig:Map_Moments}).

In the top-left panel of Figure~\ref{Fig:Map_Moments}, we show the first moment of the LSF profiles. This is in principle not a physically meaningful quantity, because the velocity zeropoint for the LSF models is arbitrary and can be freely re-defined.
Nevertheless, within the approach adopted here, the first moment provides some quite helpful insights.
We assume that the intrinsic positions of the LFC lines correspond to the locations determined in the initial Gaussian fit of the lines, expressed by the $\Cj$ values. The actual line profiles, however, might not be centered exactly on these positions.
The first moment of the inferred LSF profile is therefore a measure of how much the Gaussian-based centroid deviates from the moment-based centroid of the true profile.
In Figure~\ref{Fig:Map_Moments}, one notices a quite characteristic pattern. For the blue detector, the first moment starts to deviate quite strongly from zero towards the right edge of the orders. This deviation, however, appears only for Slice~b of both fibers and reaches a magnitude of more than $30\,\mps$.
A rather similar pattern can be seen in the bottom-left panel of Figure~\ref{Fig:Map_Moments} which shows the skewness of the profile, inferred from the third moment. Apparently, there is a quite significant asymmetry in the LSF profile that predominantly affects the right (i.e. long-wavelength) end of the spectral orders falling onto the blue detector, but exclusively for Slice~b of both fibers.
This is highly intriguing. Comparing to Figure~\ref{Fig:Comparison_ThArFP-LFC_Gauss}, one notices a strong discrepancy between slices for spectral orders recorded by the blue arm of ESPRESSO. This discrepancy increases quickly towards the red end of each spectral order and reaches nearly $30\,\mps$, which is actually a very similar behavior as we find in the first and third moment of the LSF profile.
It appears, the culprit for the discrepancy between slices has been identified, which also raises hopes that this discrepancy might disappear when taking the actual non-Gaussian LSF shape properly into account during the wavelength-calibration process.

For completeness, we also show the kurtosis of the LSF, derived from its fourth moment, in the bottom-right panel of Figure~\ref{Fig:Map_Moments}. It is a measure of how \textit{peaky} or \textit{boxy} the profiles are. In particular for the reddest part of the spectrum, there is a strong (horizontal) gradient of the kurtosis, indicating how strongly the shape of the LSF profile changes along each order.

\subsection{Re-Fitting the Calibration Spectra}
\label{Sec:ReFitting}

Characterizing the ESPRESSO LSF as done above is just by itself interesting and provides vital information about the instrument and its properties. However, improvements to the wavelength calibration can only be achieved when one manages to incorporate this knowledge into the wavelength calibration process. To do so, all following steps must make use of the LSF models, as established above. Therefore, it is necessary to re-fit the lines in the wavelength calibration spectra, taking into account the true LSF shape instead of assuming a simple Gaussian profile.
The same will later apply to the science data as well. For studies that forward model the data, e.g. by Voigt-profile fitting of absorption features like in \citet{Murphy2022}, this is readily possible. However, for other types of analysis, e.g. computation of a cross-correlation function often used in radial-velocity studies, no straight-forward solution for incorporating the LSF is immediately apparent.

For re-fitting the LFC lines, we proceed following a very similar approach as for determining the LSF. We again construct the model flux by a convolution of an intrinsic spectrum with the LSF, fully identical to Equation~\ref{Eq:Model}.
This time, however, the LSF profile is already known, but the intrinsic positions, intensities, and widths of the lines, encoded in the matrix $\A$, shall be determined. This will require a non-linear optimization process.
To simplify the problem, we fit each line individually and therefore also restrict the spectral range to just cover the one line of interest and its wings.
Following Equations~\ref{Eq:Model} and \ref{Eq:Aik_FunctionIntegration}, the model flux at pixel position $\Yi$ is therefore defined as
\begin{equation}
\Mi = \sum_k \Ij \; \int_{(k-0.5) \cdot \DV}^{(k+0.5)\cdot\DV} \f_{\Wj}\left( \Yi - \Cj - v \cdot \Disp^{-1} \right) dv \, \Pk.
\label{Eq:ModelReFit}
\end{equation}
\hl{Here, $\Pk$ describes the LSF shape, linearly interpolated between the line profiles inferred in the two adjacent blocks.
The fit parameters }$\Ij$ and $\Cj$ describe intensity and pixel location of the line to be determined. In addition, we also fit for the intrinsic line width, $\Wj$. For the LFC lines, we in principle know this, it will be very close to zero, but we still keep it as a free parameter and use it for consistency checks. The procedure is therefore also applicable to more general cases in which the intrinsic width is not known a~priori, i.e. when fitting the ThAr spectra.
We use a standard $\chi^2$ minimizer to optimize the model given in Equation~\ref{Eq:ModelReFit} and fit for the line parameters, following the likelihood definition in Equation~\ref{Eq:logL}. \hl{Formal uncertainties for the fitted parameters are estimated from the diagonal of the covariance matrix.}
There is the possibility to include an additional background component for each line. However, it was determined that this is not necessary and the a~priori determined model of the diffuse background sufficient.

When fitting lines from the FP spectra, we use for the intrinsic line profile the theoretical Airy distribution
\begin{equation}
g_\FINj(\Yi) = \frac{1}{ 1 + \frac{4\FINj^2}{\pi^2} \; \textup{sin}^2\left( \frac{\pi \; ( \Yi - \Cj )}{\mathcal{FSR}} \right) }
\end{equation}
instead of the Gaussian line profile $\f_\Wj(\Yi)$.
Here, $\FINj$ describes the finesse of the FP etalon and therefore the intrinsic width of the fitted line, quite analogous to $\Wj$ for the Gaussian line shape. In addition, the free spectral range needs to be expressed in pixels. This is given by  $\mathcal{FSR} = \frac{\textup{c}}{j}
\; \mathcal{D}^{-1}$, if $j$ is the actual physical mode number (and not only some arbitrary index). Since the modes anyway have to be correctly identified during the FP calibration process and the dispersion relation, $\Disp$, known a~priori as well (at least approximately), $\mathcal{FSR}$ acts just as a fixed conversion parameter and does not need to be fitted. Therefore, the model also has three free parameters, $\Ij$, $\Cj$, and $\FINj$, for every FP line.

\begin{figure*}
 \includegraphics[width=\linewidth]{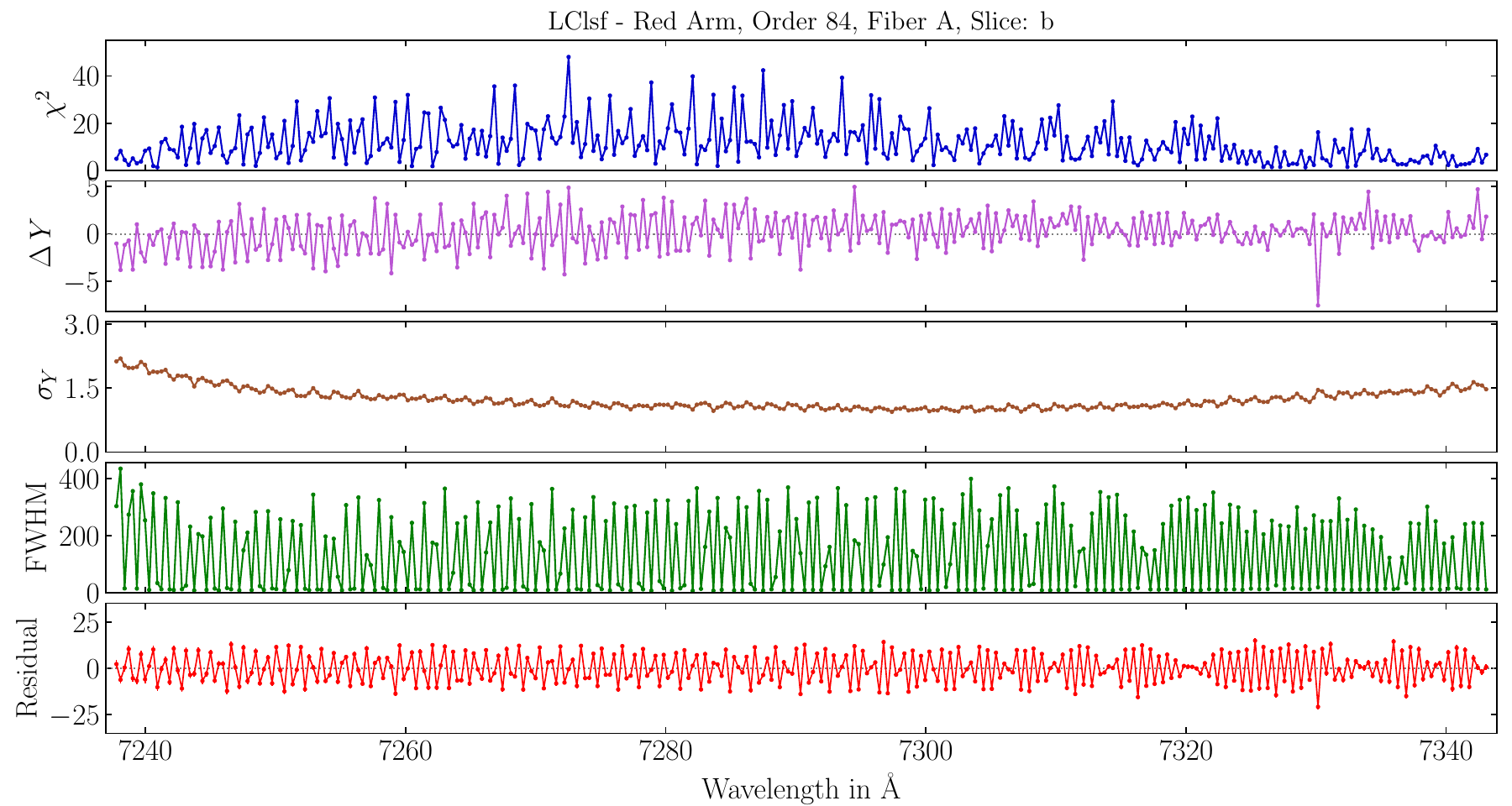}
 \caption{
  Summary of the most important quantities obtained during the re-fitting of LFC lines in Order~84, Fiber~A, Slice~b. From top to bottom, the different panels show for each individual LFC line the reduced $\chi^2$, the position difference between initial fit assuming a Gaussian line profile and final fit utilizing the non-parametric LSF model, the formal centroid uncertainty, the inferred intrinsic FWHM, and the residual between measured line position and LFC wavelength solution. All panels, except the first one, give quantities in units of $\mps$.
 }
 \label{Fig:ReFit_LFC_313}
\end{figure*}

\begin{figure}
 \includegraphics[width=\linewidth]{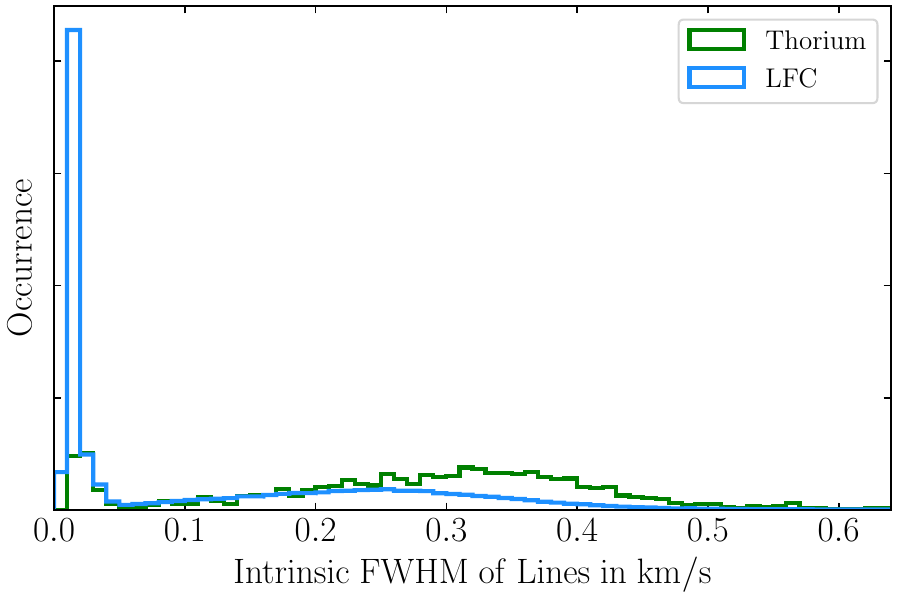}
 \caption{
  Histogram of the inferred intrinsic line widths obtained from LFC and ThAr spectra, combining data from all orders, fibers, and slices.
 }
 \label{Fig:HistogramLineWidths}
\end{figure}

Figure~\ref{Fig:ReFit_LFC_313} shows the results of the LFC line re-fitting process for one selected spectral order. Using the determined LSF shape does lead to a much better, but still not perfect match between data and model, and thus $\chi^2$ values substantially larger than unity. This is not particularly surprising, given that already during the construction of the LSF model significant residuals appeared (see Figure~\ref{Fig:Flux_ModelData}).
We define the line centroid difference, $\Delta{}Y$, as the difference between fits assuming either a Gaussian or an empirical LSF model.
By construction, it is close to zero for the LFC lines, since the initial line positions used to construct the LSF model were based on Gaussian fits.
A striking feature in Figure~\ref{Fig:ReFit_LFC_313}, however, is the fitted intrinsic width of the LFC lines. This is expected to be very close to zero. The construction of the LSF profile assumed a width of $6\,\cmps$, and the subsequent fit of the lines with exactly this LSF model should return (within the uncertainties) the same value. However, one observes that for only about half of the lines the determined intrinsic width is indeed negligible, while for the other half one finds $\approx250\,\mps$. Here, \textit{narrow} and \textit{wide} lines follow in quick succession with nearly every other line having a non-vanishing width. This, again, reflects the systematics that are obviously present in the data and already appeared in Figure~\ref{Fig:Flux_ModelData}. These can by construction not be captured by the LSF model, because it assumes that the LSF shape is constant within each 577~pixel block  and in consequence reflects the average of multiple \textit{narrow} and \textit{wide} lines.
It remains an open question what causes this modulation in the apparent line width. The effect clearly resembles the \textit{beat pattern} described in \citet{Schmidt2021} and shown for reference in the bottom panel of Figure~\ref{Fig:ReFit_LFC_313}. The cause could in principle be related to some adverse optical effects, i.e. irregularities in the ruling of the grating, and therefore be real, or caused by the pixelization of the data and therefore stem from inaccuracies in the spectral extraction process.
Spectral extraction is performed using the algorithm presented in \citet{Zechmeister2014}, which assumes--like nearly all extraction schemes adopted in astronomy--that the two-dimensional instrumental profile is a separable function w.r.t. the two coordinates. Already in \citet{Schmidt2021}, it has been demonstrated that this leads to some inaccuracies in the extraction process. However, it still remains unclear how large the actual impact on the one-dimensional spectra is and whether the observed \textit{beat pattern} is related to it or not. A definitive answer will require a dedicated study and probably the implementation of a substantially more sophisticated extraction algorithm, like the one presented by \cite{ABolton2010}.

A histogram of the inferred intrinsic line widths is presented in Figure~\ref{Fig:HistogramLineWidths}. Again, the described bimodality in the apparent line width is obvious. In addition, it highlights that the thorium lines are actually intrinsically wider than the LFC lines. The bimodality in the line widths obviously affects them as well, but less lines exhibit vanishing widths and the bulk of the distribution is located at roughly $100\,\mps$ larger width.

\begin{figure*}
 \includegraphics[width=\linewidth]{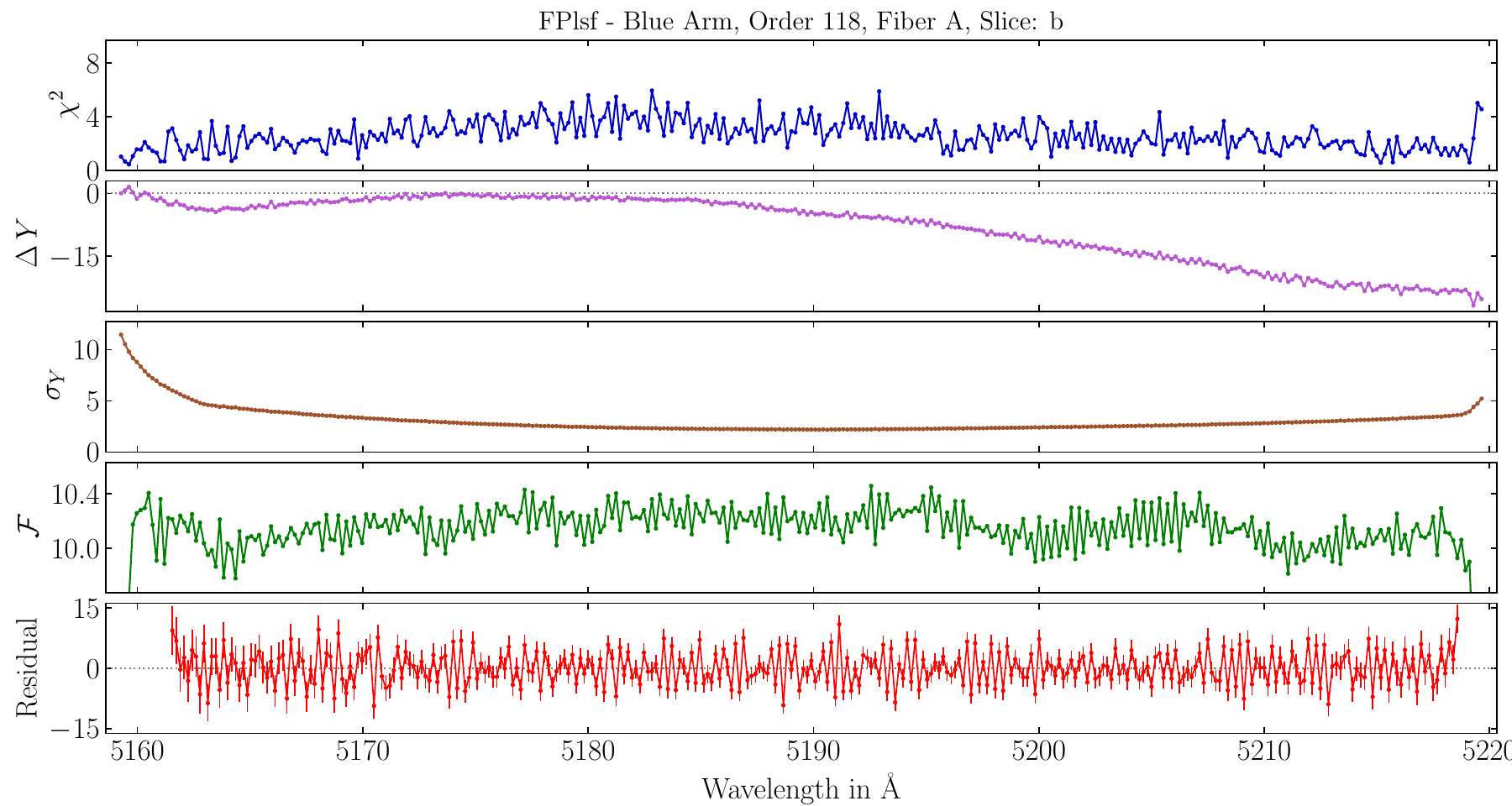}
 \caption{
  Summary for the re-fitting of the FP lines in Order~118, Fiber~A, Slice~b.
  Quantities are identical to Figure~\ref{Fig:ReFit_LFC_313}, except that the panel second from the bottom shows the finesse of the FP. Also, the  bottom most panel displays the residuals of the FP lines w.r.t. the ThAr/FP wavelength solution.
 }
 \label{Fig:ReFit_FP_173}
\end{figure*}

A similar plot to Figure~\ref{Fig:ReFit_LFC_313} can also be made for the re-fit of the FP lines. This is shown in Figure~\ref{Fig:ReFit_FP_173}.
The fit of the FP lines is actually formally better than for the LFC and the $\chi^2$ lower.
Now, the line shift $\Delta{}Y$ exhibits a clear pattern, dropping systematically to $-20\,\mps$ towards the right end of that order.
We stress again that the absolute value for this offset is fully arbitrary, just related to the definition of the LSF, and has by itself no physical meaning. Relevant is only the observed difference between various kinds of spectral sources. Obviously, using the non-Gaussian LSF model instead of a simple Gaussian profile has different impacts for LFC and FP spectra. Most likely, the effect for the FP is actually smaller, because the resolved nature of the relatively wide FP lines dampens the sensitivity to the LSF shape. However, in our scheme, the LFC lines are defined to be the reference and their offset is thus by construction close to zero. In consequence, any significant shift must appear for the FP lines. The clear $\Delta{}Y$ pattern shown in Figure~\ref{Fig:ReFit_FP_173} already indicates that the comparison between ThAr/FP and LFC solution (Figure~\ref{Fig:Comparison_ThArFP-LFC_Gauss}) will look substantially different when using the more correct LSF model.

\begin{figure}
 \includegraphics[width=\linewidth]{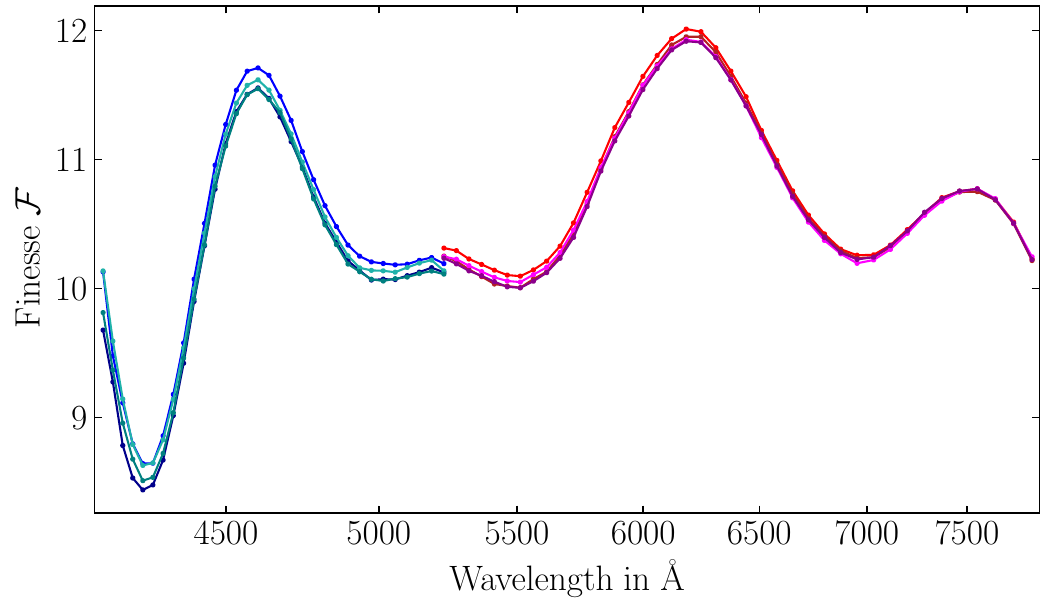}
 \caption{
  Inferred finesse of the ESPRESSO Fabry-P\'erot device over the full wavelength range covered by the LFC. Colors indicate measurements obtained from different fibers and slices.
 }
 \label{Fig:Finesse}
\end{figure}

Similarly to the LFC fit, we also find a clear semi-periodic modulation of the line width, here expressed as finesse $\mathcal{F}$. The ESPRESSO FP has a relatively moderate design finesse ($\mathcal{F}\approx12$, \citealt{Schmidt2022}) and therefore produces lines with an intrinsic FWHM of about $200\,\textup{MHz}$, corresponding to widths slightly below $1\,\kmps$, which is less than half of the spectral resolution, but already quite noticeable and vastly different from the narrow LFC or thorium lines. Still, the issue regarding the apparent line-width modulation also exists for the FP, indicating that this is not related to just the LFC.
Apart from this issue, we receive, as a side-product, a full characterization of the FP finesse.
Measuring this with the spectrograph itself is only reliably possible due to the availability of an accurate LFC model.
The determined finesse over the full spectral range is presented in Figure~\ref{Fig:Finesse}. Unfortunately, no detailed characterization of the ESPRESSO FP was performed before it was shipped to the observatory and thus, no direct comparison with a laboratory measurement is possible. Nevertheless, Figure~\ref{Fig:Finesse} highlighting that the inferred finesse is fairly consistent for the different fibers, slices, and even across spectrograph arms, building substantial confidence in the accuracy of the measurement. This showcases that modeling of the instrument LSF indeed allows to accurately measure spectral structures at the sub-resolution scale.

\subsection{Constructing the Wavelength Solutions}
\label{Sec:WavelengthSolutions}

\begin{figure*}
 \includegraphics[width=\linewidth]{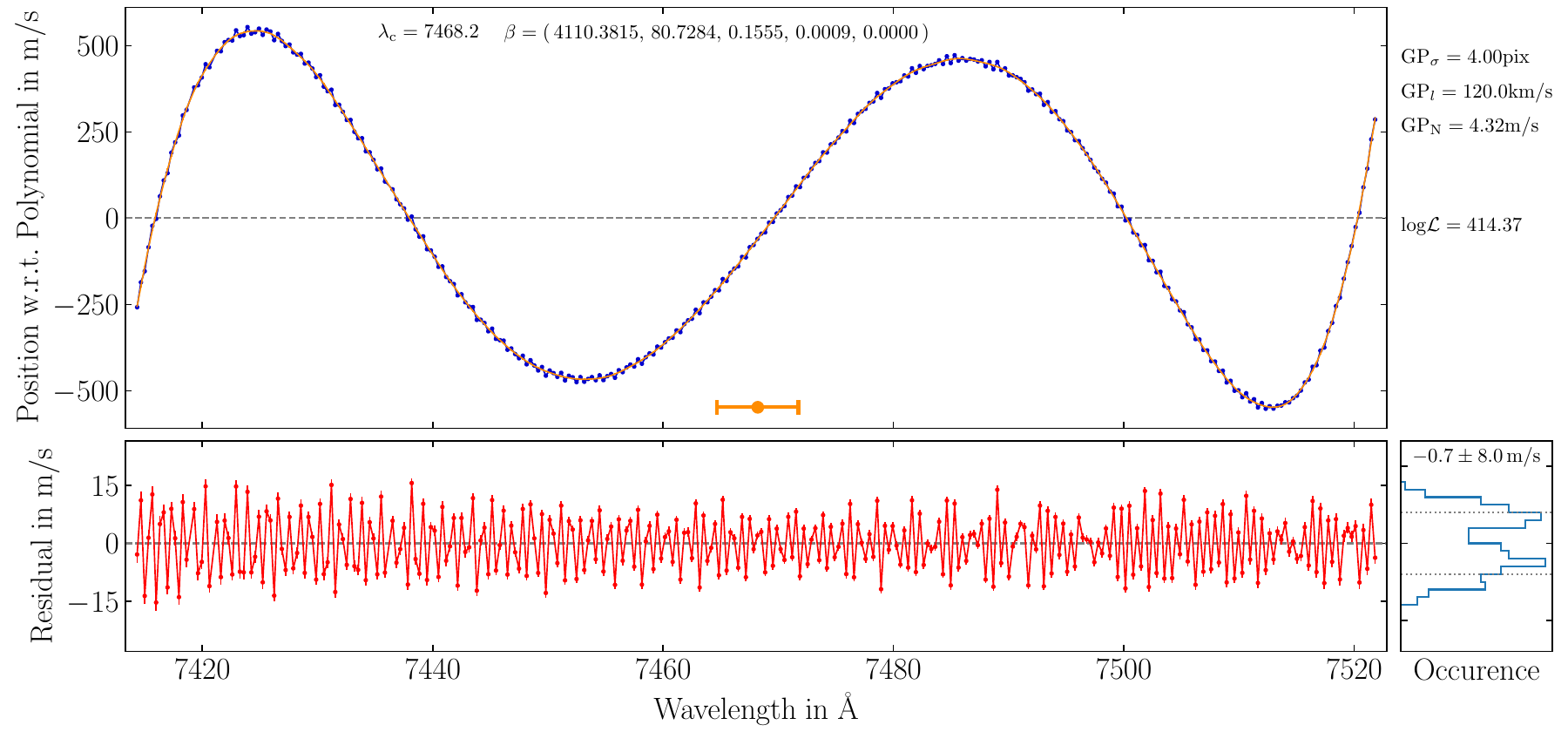}
 \caption{
  Visualization of the intra-order wavelength solution for Order~82, Fiber~A, Slice~a. The top panel shows the determined positions of the LFC lines (blue) and the derived LFC wavelength solution (orange), both w.r.t. a polynomial that was subtracted to reduce the dynamic range. The defining parameters for the polynomial, following Equation~\ref{Eq:Polynomial}, are given at the top. The bottom panel shows the residuals of the LFC lines w.r.t. the LFC wavelength solution. Formal uncertainties are indicated by error bars. The right panel contains the corresponding histogram. The size of the kernel utilized for constructing the wavelength solution is indicated at the bottom of the large panel with an orange, horizontal bar.
 }
 \label{Fig:IntraOrderWavelengthSolution}
\end{figure*}

After re-fitting all lines in the different types of wavelength calibration exposures, one has to construct wavelength solutions from them. Here, we closely follow the procedure described in detail in \citet{Schmidt2021}.
LFC modes, $j$, are identified using some a~priori available wavelength solution. This is readily possible due to the high stability of ESPRESSO. Following the fundamental LFC relation, the true LFC line frequencies are therefore
\begin{equation}
 \nu_j  = \nu_0 + j \cdot \nu_\textup{sep} .
\end{equation}
Here, $\nu_0$ and $\nu_\textup{sep}$ are the offset frequency and separation of the comb lines, both actively controlled by the LFC system and listed in the \texttt{fits} headers.
The FP effective gap size $D_\textup{eff}(\lambda)$ is calibrated using the ThAr spectra, following the procedure described in \citet{Cersullo2019} and \citet{Schmidt2021}. Therefore, also the true wavelengths of each FP line are known.
The remaining task is to construct a continuous wavelength solution from the determined wavelengths and pixel positions of the $\gtrsim200$  LSF or FP lines for each spectral order. In \citet{Schmidt2021} we used a modification of the \citet{SavitzkyGolay1964} filter for this. Here, we instead adopt an approach based on Gaussian process regression. Since we are dealing here, in contrast to the LSF determination, with a quite ordinary regression problem, one can use the standard Gaussian process regression methods, described with great detail in the literature, e.g. in \citet{Rasmussen2006}.

The calculation of the wavelength solution for each order is phrased in a backward manner, i.e. the data pairs $(\Cj | \lambda_j )$ extracted from the FP or LFC spectra are used to construct a solution of the form $\Y(\boldsymbol{\lambda})$. This then still has to be inverted to get the wavelength as function of pixel position.
Formulating the GP as the inverse problem, i.e. directly computing $\boldsymbol{\lambda}(\Y)$, would be more elegant but also comes with some complications. In particular, we want to specify the correlation length of the kernel in physically meaningful units, i.e. in $\kmps$. This would require to compute some sort of preliminary wavelength solution to convert the specified kernel from velocity to pixel coordinates. We therefore decide to first establish $\Y(\boldsymbol{\lambda})$, evaluate this function on a dense wavelength grid, and then do the inversion using cubic spline interpolation.

Gaussian processes are often formulated with a vanishing prior mean. This is in stark contrast to the a priori knowledge we have about the wavelength solution: The pixel positions as function of wavelength are not expected to be all located around zero but will monotonically increase with wavelength and should, as a very rough approximation, follow a linear relation. We therefore adopt the approach of a Gaussian process with explicit basis functions, as described in \citet[][Section 2.7]{Rasmussen2006}.
Here, we incorporate a set of polynomial functions, which are supposed to capture the global trend of the wavelength solution and ensure a better asymptotic behavior towards the order edges. The classical Gaussian process component is then supposed to accurately capture the remaining variation of the wavelength solution around the polynomial.
Following the notation in \citet{Rasmussen2006}, our model is a combination of a classical Gaussian process, $\mathcal{GP}(\boldsymbol{\lambda})$, with zero mean and covariance $K(\boldsymbol{\lambda},\boldsymbol{\lambda'})$, and a set of polynomial functions contained in $\boldsymbol{h}(\lambda)=(1,\lambda,\lambda^2,\dots)$ with coefficients $\boldsymbol{\beta}$. Both together can be written as
\begin{equation}
 \Y(\boldsymbol{\lambda}) = \mathcal{GP}[\,\boldsymbol{\lambda}\,|\,0,K(\boldsymbol{\lambda},\boldsymbol{\lambda'})\,] + \boldsymbol{h}(\boldsymbol{\lambda})^\top \; \boldsymbol{\beta} .
 \label{Eq:GP_Poly}
\end{equation}
For the covariance, we adopt a simple squared-exponential kernel
\begin{equation}
 K(\lambda,\lambda') = \frac{ {\sigma_{\lambda}}^2 }{ \sqrt{ 2 \pi {L_{\lambda}}^2 } } \; \textup{exp}\left[ -\frac{1}{2} \frac{ \textup{c}^2 \left( \frac{\lambda \: - \: \lambda'}{0.5\lambda \: + \: 0.5\lambda'} \right)^2 }{ {L_{\lambda}}^2 } \right]
\end{equation}
in which the correlation length, $L_{\lambda}$, is expressed in velocity units.
To improve the numerical stability, we parameterize the polynomials as $\boldsymbol{h}(\lambda - \lambda_\textup{c})$, i.e. relative to the central wavelength, $\lambda_\textup{c}$, of any given spectral order. The remaining priors for the polynomial part are assumed to be uninformative. The model described in Equation~\ref{Eq:GP_Poly} is again a Gaussian process for which the posterior probability distribution, i.e. the posterior mean, $\overline{\Y}(\boldsymbol{\lambda})$, and the corresponding covariance, can be computed analytically. In addition, one can also calculate the posterior mean for the coefficients of the polynomials, $\overline{\beta}$, which provides a quite intelligible description of the polynomial part of the solution.

Figure~\ref{Fig:IntraOrderWavelengthSolution} visualizes the intra-order wavelength solution constructed from the LFC spectrum of one selected spectral order.
The top panel shows the data, $(\Cj | \lambda_j )$, as well as the fitted model, $\overline{\Y}(\boldsymbol{\lambda})$.
Since the overall dynamic range would be far too large to show in a meaningful plot, the polynomial component of the wavelength solution, given by
\begin{equation}
 Y^\textup{pol}_i(\lambda_i) = \sum_{k=0}^N \; \overline{\beta}_k \; (\,\lambda_i - \lambda_\textup{c}\,)^k ,
\label{Eq:Polynomial}
\end{equation}
was subtracted from data and model. The figure therefore only shows the $\mathcal{GP}$ part from Equation~\ref{Eq:GP_Poly}, but $\lambda_\textup{c}$ and $\overline{\boldsymbol{\beta}}$ are stated in the figure.
For the polynomial component, a function of fourth order was chosen. We also investigated polynomial models of order three and five, but all of them perform together with the remaining Gaussian process equally well. For the correlation length, we use $L_{\lambda}=120\,\kmps$ (indicated in Figure~\ref{Fig:IntraOrderWavelengthSolution} with a horizontal bar) and $\sigma_{\lambda}=4\,\textup{pix}$.

The bottom panel of Figure~\ref{Fig:IntraOrderWavelengthSolution} shows the residuals of the data w.r.t. the model. Obviously, these are quite substantial and exceed the formal uncertainties by far. The residuals also exhibit a characteristic pattern with very fast oscillations.
This effect was already discovered in \citet{Schmidt2021}, labeled as \textit{beat pattern noise}, and discussed in detail, considering  numerous possible or impossible causes. Since then, there has been little progress in pinpointing the origin of the effect and Figure~\ref{Fig:IntraOrderWavelengthSolution} demonstrates that it is still there at a comparable strength, even when properly modeling the LSF.
This Figure also shows that, as already stated in \citet{Schmidt2021}, the residuals do not depend on the particular algorithm used to construct the wavelength solution. The scatter of the measured line positions becomes visible by eye already in the top panel of Figure~\ref{Fig:IntraOrderWavelengthSolution}, where only a polynomial was subtracted. Every smooth wavelength solution will lead to more or less the same residuals. Therefore, the \citet{SavitzkyGolay1964} filter used in \citet{Schmidt2021} basically performs as good as the approach adopted here. The main advantage of the Gaussian process is that it actually provides formal uncertainty estimates for the wavelength solution. Nevertheless, the \textit{beat pattern noise} remains and of course dominates the uncertainties in the wavelength solution, in particular at redder wavelengths.

The final task is to invert the wavelength solution to provide a relation $\boldsymbol{\lambda}(\Y)$. For this, the model, $\overline{\Y}(\boldsymbol{\lambda})$, established above, is evaluated on a dense grid of wavelengths (i.e. at 1200 positions per spectral order) and a simple cubic spline interpolation is then used to invert and interpolate these samples to the pixel grid. This can be done without complications, given that the samples are closely spaced and the underlying function is--by construction--anyway smooth on the scales relevant for the interpolation.

\subsection{Comparison of ThAr/FP to LFC Wavelength Solution}

\begin{figure*}
 \includegraphics[width=\linewidth]{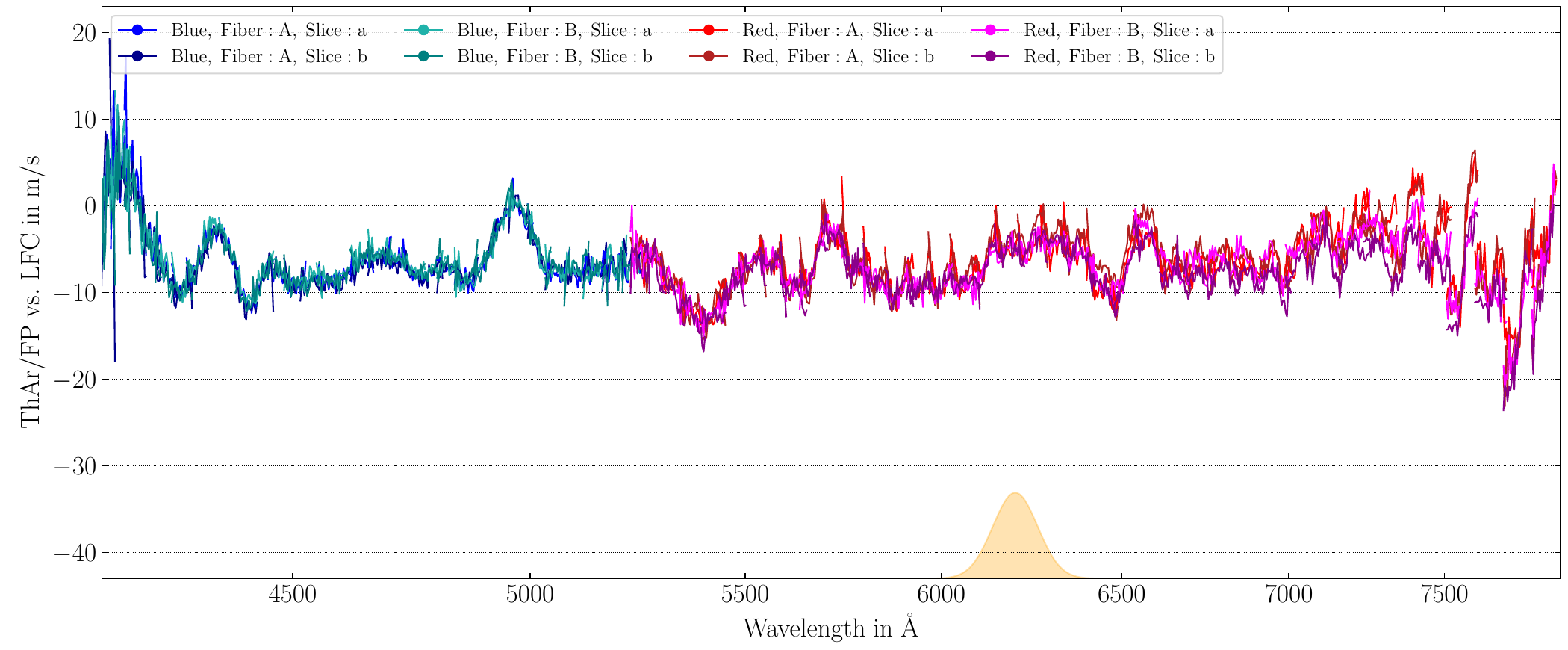}
 \caption{
  Comparison between ThAr/FP and LFC wavelength solution, constructed using the empirical, non-parametric LSF model. The data itself is identical to Figure~\ref{Fig:Comparison_ThArFP-LFC_Gauss}. Colors represent calibrations obtained for different fibers and slices. The orange Gaussian at the bottom represents the smoothing kernel used for calibrating the FP effective gap size.
 }
 \label{Fig:Comparison_ThArFP-LSF_LSF}
\end{figure*}

After constructing the joint ThAr/FP wavelength solution and a fully independent calibration based on the LFC spectra, one can compare both of them, as we have done in Figure~\ref{Fig:Comparison_ThArFP-LFC_Gauss}, but this time with proper treatment of the ESPRESSO LSF. The result is shown in Figure~\ref{Fig:Comparison_ThArFP-LSF_LSF}.
Obviously, this new comparison looks a lot cleaner, more consistent, and with less systematics than the old one based on Gaussian line fits. It therefore marks a major step forward.

Most strikingly, the discrepancy between slices in the blue arm of ESPRESSO has disappeared. With a proper modeling of the LSF, both fibers and slices now provide fully consistent information. An improvement like this has already been expected, given how different the asymmetry of the LSF is in the two slices (Figure~\ref{Fig:Map_Moments}), but Figure~\ref{Fig:Comparison_ThArFP-LSF_LSF} demonstrates that incorporating an accurate LSF profile actually solves this problem.
Nevertheless, ThAr/FP and LFC wavelength solutions still do not fully agree in an absolute sense. There remains an offset of about $7\,\mps$, less than in Figure~\ref{Fig:Comparison_ThArFP-LFC_Gauss} but clearly present, and some modulation at the $\pm8\,\mps$ level.
The crucial and most fundamental aspect in this context, however, is internal consistency, i.e. that all four traces and overlapping orders provide compatible results, and that there are no strong intra-order modulation.
This now has been achieved, at least for wavelengths $\lambda\lesssim7000\,\AAA$, and indicates that the data-processing and spectrograph-related inaccuracies can mostly be eliminated by a careful modeling of the LSF.

The remaining discrepancy between ThAr/FP and LFC solution might in fact be related to the ThAr spectra, which provide all the absolute wavelength information for the ThAr/FP solution.
Here, there could be blends, misidentifications, or contaminations of lines, but also the catalog of laboratory wavelength, taken from \citet{Redman2014}, could be the origin of some inaccuracies. For instance, Fourier-transform spectrometer, as used by \citet{Redman2014},
have notorious difficulties in determining the absolute wavenumber scale due to unavoidable non-common-path aberrations between the tracking laser and the to-be-measured spectrum. This might lead to a global velocity offset. The formal uncertainty given by \citet{Redman2014} is only about $0.5\,\mps$, but there could be additional systematics at the few~$\mps$ level.
Other issues could be related to individual thorium lines.
Figure~\ref{Fig:Comparison_ThArFP-LSF_LSF} exhibits some pronounced deviations from the global trend at e.g. $4350$, $4960$, and $5400\,\AAA$.
These structures have the size and shape of the smoothing kernel (indicated in orange in Figure~\ref{Fig:Comparison_ThArFP-LSF_LSF}), which was used when calibrating the FP effective gap size (\,$D_\textup{eff}(\lambda)$, see \citealt{Schmidt2021} for details).
In principle, individual outliers among the thorium lines could cause such patterns.
With the improved and now internally consistent wavelength calibration just established, it would make sense to have a closer look at the ThAr spectra, the selection of lines, their laboratory wavelengths and so forth. There is probably a significant potential for improvements. For instance, using an accurate LSF model when fitting the thorium lines allows to judge much better whether a line is actually isolated and reliable or in fact blended or otherwise problematic. However, this is a separate problem, requires a study on its own, and will be dealt with somewhere else.

The focus here is therefore mostly on the consistency of the wavelength solutions. Although massively improved, there do remain some issues.
For longer wavelengths ($\gtrsim6000\,\AAA$), the comparison between ThAr/FP and LFC solution becomes increasingly noisy and more affected by systematics. This might be due to different reasons:
The amplitude of the \textit{beat pattern} shown in Figure~\ref{Fig:IntraOrderWavelengthSolution} and described in \citet{Schmidt2021} increases with wavelength. Therefore, each individual line measurement is affected by a (pseudo random) noise of up to $\pm8\,\mps$ which dominates by far over the stochastic photon noise and propagates into the derived wavelength solutions, leading to larger uncertainties in redder spectral orders.
The \textit{beat pattern} effect also affects the individual thorium line measurements used to calibrate the FP. In addition, there are just fewer usable thorium lines at longer wavelengths. Thus, the FP effective gap size is just not as well constrained at long wavelengths as it is for short ones. This might account for some of the large-scale structure in Figure~\ref{Fig:Comparison_ThArFP-LSF_LSF}.
However, fluctuations on the scales of individual spectral orders, noticeable in the red half of the spectral range, can not be related to the ThAr frames, because the kernel used during the FP calibration is much wider than one spectral order.

One important question is of course whether all structure in the instrumental LSF is accurately captured by our model or if there remain additional components that are unmodeled. The blue and red arm of ESPRESSO have separate cross-dispersers, cameras, and detectors. Their optical properties are therefore different, which also leads to unequal LSFs.
The most prominent feature for the blue detector in Figure~\ref{Fig:Map_Moments} is the different line asymmetry for the two slices (first and third moment).
The strongest effect for the red detector, on the other hand, is the change in kurtosis (\textit{boxyness}, fourth moment) along the individual orders, in particular at the longest wavelengths. In case our LSF model would not be flexible enough to fully capture this rapid change of line profile, one could expect residual intra-order structure in the wavelength comparison.
We therefore varied several hyperparameters of our analysis scheme (listed in Table~\ref{Tab:Hyperparameters}), i.e. separating each spectral order into more and finer blocks, therefore allowing faster variations of the LSF profile along order, or choosing a shorter correlation length to make the LSF model in general more flexible.
The effect of this was quite apparent for the LSF models and in accordance with our expectations, i.e. the inferred profiles became more accurate but also more noisy.
Nevertheless, this had no significant impact on the comparison of wavelength solutions. We therefore conclude that the LSF model is currently not the limiting factor for the wavelength calibration accuracy.

We also tried to optimize the construction of the wavelength solutions described in Section~\ref{Sec:WavelengthSolutions}. This actually lead to some improvements, e.g. by rejection of outliers. Also, we removed some parts at the beginning and end of each spectral order which seem to be affected by vignetting and caused significant edge effects. A bit of this is still present in Figure~\ref{Fig:Comparison_ThArFP-LSF_LSF}, but it is not the dominant effect.
Inspecting the residuals in Figure~\ref{Fig:IntraOrderWavelengthSolution} also reveals that no further multi-$\mps$ improvements can be achieved here. The individual line measurements scatter quite significantly around the wavelength solution, but they do this in a symmetric fashion and the model does capture the mean quite well. We therefore do not see any practical approach to substantially improve the wavelength calibration accuracy within the scheme outlined in this work, at least as long as the \textit{beat pattern} prevails.

Some particular large systematics appear for the five reddest spectral orders at $\lambda>7000\,\AAA$.
Here, it appears as if the individual orders were split in the middle and the left halves exhibit a more negative offset between ThAr/FP and LFC solution while the right ones show a less negative or even positive difference. In consequence, the (marginally) overlapping edges of consecutive orders provide no consistent difference between wavelength solutions. Such a behavior is puzzling. We note that the orders are actually split in the center and read out to both edges of the detector. Therefore, charge-transfer inefficiencies \citep[e.g.][]{Goudfrooij2006, Bouchy2009, Blake2017, Blackman2020} that affect FP and LFC lines differently could in principle cause such an effect. We also note that the intensity of the FP lines drops substantially for $\lambda>7000\,\AAA$ (see Figure~\ref{Fig:Flux_LCFP}). However, no such intra-order discrepancies appear in other spectral region with strong fluctuations of either LFC or FP intensities, like in the region $5000 < \lambda < 5500\,\AAA$.
Also, one would expect that charge-transfer inefficiencies would have a vanishing effect for the edges of the orders, which are closest to the output amplifiers. Still, the overlapping regions of neighboring order edges exhibit significant discrepancies. The observed pattern therefore seems not fully consistent with the expectation for charge-transfer inefficiencies and it therefore remains unclear whether these actually provide a plausible explanation for the observed effect. In any case, a thorough and comprehensive study of the effects related to charge-transfer inefficiencies in ESPRESSO data has to be done at some point.

Overall, however, Figure~\ref{Fig:Comparison_ThArFP-LSF_LSF} demonstrates a massive improvement in the wavelength calibration accuracy, reducing the discrepancies between ThAr/FP and LFC wavelength solutions from approximately $50\,\mps$ to about $10\,\mps$ (peak-to-valley, and excluding the few reddest orders). Most importantly, it demonstrates that accurately modeling the LSF nearly completely eliminates spectrograph-related inaccuracies, i.e. discrepancies between fibers and slices and intra-order modulations, and reduces their amplitude to below a couple $\mps$ over the majority of the wavelength range.

\section{Summary and Conclusions}

In this study, we have presented a method to substantially improve the ESPRESSO wavelength calibration accuracy, specifically by accurate modeling of the instrumental LSF.
A previous analysis of the ESPRESSO wavelength calibration accuracy, carried out by \citet{Schmidt2021}, demonstrated that substantial discrepancies exist between the ThAr/FP and LFC wavelength solutions when fitting all lines with simple Gaussian functions. From the characteristics of these discrepancies, one could conclude that at least the most dominant ones must be related to the spectrograph and data processing (Figure~\ref{Fig:Comparison_ThArFP-LFC_Gauss}).
A probable cause, in particular for the intra-order modulations, was a deviation of the instrumental LSF from the assumed Gaussian shape.
In this study, we therefore construct an accurate, non-parametric model of the ESPRESSO LSF. For this, we make use of the LFC spectra, which provide a plethora of narrow, unresolved lines and thereby reveal the instrumental LSF, and adopt a Gaussian-process related algorithm to fit for the intrinsic LSF, closely following the approach presented by \citet{Hirano2020}.

The determined LSF shapes are clearly non-Gaussian, asymmetric, different for the two slices and fibers, and vary significantly across the detectors (Figure~\ref{Fig:LSF1+2}, \ref{Fig:LSF_Profiles_169+329}, \ref{Fig:Map_FWHM}, and \ref{Fig:Map_Moments}).
Re-fitting the lines in the calibration spectra using the determined LSF model leads to significant shifts of the determined positions (mostly for the FP spectra since the LFC lines are defined as reference for the LSF model).
In consequence, also the constructed ThAr/FP and LFC wavelength solutions change.
Figure~\ref{Fig:Comparison_ThArFP-LSF_LSF} demonstrates that this dramatically improves their consistency and therefore the ESPRESSO wavelength calibration accuracy in general. The overall discrepancy is reduced from $50\,\mps$ to just slightly more than $10\,\mps$ (peak-to-valley). The even more important aspect is, that now the different fibers and slices provide fully consistent results, with typical scatter smaller than a couple $\mps$. Also the intra-order modulations are drastically reduced.

However, some issues remain. For wavelength $\lambda > 7000\,\AAA$, discrepancies between the fibers and slices become quite noticeable. We speculate that this might be related to charge-transfer inefficiencies, but the true cause is unknown.
Furthermore, the \textit{beat pattern}, a fast, quasi-periodic displacement of LFC or FP lines w.r.t. the corresponding smooth wavelength solution, already reported in \citet{Schmidt2021}, still prevails even when using the correct LSF model (Figure~\ref{Fig:IntraOrderWavelengthSolution}). Now, we see in addition a very similar fast modulation of the apparent intrinsic line width for LFC and FP spectra (Figure~\ref{Fig:ReFit_LFC_313} and \ref{Fig:ReFit_FP_173}, respectively). Together with the systematic residuals appearing during the determination of the LSF (Figure~\ref{Fig:Flux_ModelData}), this provides evidence that--at least in the extracted one-dimensional spectra--neighboring lines do not exhibit identical line shapes but instead are subject to a very fast (period of approx two LFC or FP lines, corresponding to $\approx36\,\textup{GHz}$) modulation in the (apparent) line shape. The reason for this remains unclear.
Similar, but not necessarily identical effects have been observed in spectra taken with other spectrographs.
Although this effect can to a certain degree be mitigated by smoothing over multiple lines, it now represents the dominant systematic on the level of individual lines.
A dedicated study aimed at better characterizing this \textit{beat pattern}, reviewing the occurrence for different instruments, pinpointing its origin, and--if possible--finding a cure for it will therefore be a top priority for the near future. Extremely helpful for this endeavor would be a tunable laser frequency comb, allowing to measure the effect with a much finer sampling than permitted by the separation of individual LFC lines. Unfortunately, demonstration of the tuning capability of the ESPRESSO LFC is still pending.
Another issue is the global offset and some aspects of the medium-scale structure visible in Figure~\ref{Fig:Comparison_ThArFP-LSF_LSF}, which are most-likely related to the ThAr calibration. Here, significant further improvements should be possible. Relying on the accurate LSF model now available, it will be possible to judge with much greater certainty whether individual thorium lines are in fact isolated and unblended or instead pairs, contaminated, or otherwise unsuitable for the calibration process. With some effort, it should therefore be possible to obtain a better line selection and thereby remove the medium-scale distortions visible in Figure~\ref{Fig:Comparison_ThArFP-LSF_LSF}.
In addition, it might also be possible to obtain a better, in particular more accurately calibrated, laboratory measurement of the ThAr spectrum.
Thus, there remains room for improvement, but also, for at least some aspects, a clear path how to achieve this. Nevertheless, the presented characterization and incorporation of the non-Gaussian LSF in the calibration process showcases a remarkable step forward in terms of ESPRESSO wavelength calibration accuracy.

\hl{
The most-suitable approach for incorporating the ESPRESSO LSF in  the analysis of actual science observations depends strongly on the type of analysis that is performed on the data. In a forward-modeling process, where an intrinsic spectrum is convolved with the LSF to described the observed data, this is readily possible.
The typical studies focusing on the fine-structure constant \citep[e.g.][]{Webb1999, King2012, Murphy2022} follow this approach and model the absorption features by a composition of analytic Voigt profiles. Incorporating the ESPRESSO LSF model in such a type of analysis is straight forward and not substantially different from our fit of the FP data (Section~\ref{Sec:ReFitting}), where we as well assumed an analytic model of the intrinsic line shape, convolved it with the determined LSF, and then optimized the model-defining parameters to achieve a good fit to the data. The same can be done for a model composed of multiple Voigt profiles. The computational expense might become relevant, but there is no inherent complication in using the ESPRESSO LSF model for a fine-structure constant analysis.
}

\hl{
For studies that do no follow a forward-modeling approach, this is far more difficult.
For instance, many RV studies of exoplanets use a form of cross-correlation analysis. Here, no representation of the intrinsic spectrum is available and would first have to be constructed somehow. \citet{Hirano2020} describe an approach to construct a \textit{de-convolved} template of the target stars. This, however, is not trivial and its feasibility might depend on the individual science cases.
}

\hl{
More generally, we stress that the instrumental LSF varies strongly across the detector. Under these circumstances, co-addition of different spectral orders and slices to one continuous spectrum is thus in general not possible anymore without loss of accuracy. To preserve the full information content and be able to take the LSF properly into account, each extracted trace (spectral order, fiber, slice) has to be analyzed separately and in conjunction with its corresponding LSF model.
}

\hl{While this study focuses exclusively on \textit{accuracy} aspects, it is also worth to briefly touch upon the \textit{stability} of the wavelength solution.} Here, substantially tighter requirements apply, e.g. a goal of $10\,\cmps$ RV stability for ESPRESSO, but on the other hand, the comparison is only relative, between different measurements taken with the same instrument \hl{at different times}. It was therefore often assumed that the exact shape of the instrumental LSF does not matter, as long as it remains stable.
Preliminary studies determined that this is not entirely correct. Changes in the spectra itself, i.e. variations in the S/N ratio or relative contribution of various components like back- and foregrounds, will indeed lead to spurious RV shifts that can be substantially reduced by adopting a correct model for the LSF.
Therefore, achieving the most extreme RV precision likely requires as well a proper treatment of the instrumental LSF.
A dedicated study regarding this aspect is ongoing and will be presented in the near future.

It becomes clear that incorporating a detailed model of the instrumental LSF is essential to achieve \textit{accurate} wavelength measurements and probably also important for \textit{stable} RV measurements.
A key difficulty here is to characterize the instrumental LSF with sufficient detail. This requires appropriate calibration spectra that actually reveal the LSF. Currently, the LFC is the best, but also only, available calibration source that provides suitable spectra.
However, our study also revealed the limitation of the adopted approach. Determining the LSF on smaller blocks, with higher precision and lower noise, would require more informative calibration data. These could come in the form of a tunable LFC, that allows to sweep the lines over the detector and obtain a more detailed and finer sampled measurement of the LSF.
Alternatively, given the technical difficulties involved in such a system, one could also image to develop a dedicated high-finesse FP device, just for LSF determination. Here, the most important aspect would be that the intrinsic line shape is accurately known a~priori (required in Equation~\ref{Eq:Aik_FunctionIntegration}). On the other hand, and in contrast to the existing FPs used for wavelength calibration, no particularly high level of stability would be needed when only the LSF shape shall be measured. Therefore, tuning of such a FP might become feasible. Which concept might turn out to be the most practical one remains to be seen, but there clearly exists a need for calibration sources that allow a high-quality determination of the LSF.
Accurately modeling the instrumental LSF and incorporating it in the data analysis will be essential for further pushing the limits of accurate and stable high-resolution spectroscopy. Quite substantial improvements could be demonstrated in this study, but achieving in particular the extremely ambitious goals of ANDES \citep[e.g.][]{Marconi2022, Martins2023} will require further development of algorithms and methods, but also the introduction of dedicated LSF calibration sources, e.g. tunable LFCs or special high-finesse FP calibrators.

\section*{Acknowledgements}

We thank the anonymous reviewer for helpful comments.
This research has made use of Astropy, a community-developed core Python package for Astronomy \citep{Astropy2013,Astropy2018,Astropy2022}, and Matplotlib \citep{Hunter2007}.
TMS  acknowledgment the support from the SNF synergia grant CRSII5-193689 (BLUVES).
This work has been carried out within the framework of the National Centre of Competence in Research PlanetS supported by the Swiss National Science Foundation.

\section*{Data Availability}

All data used for this study is publicly available from the ESO archive facility: \url{http://archive.eso.org/cms/data-portal.html}.


\bibliographystyle{mnras}
\bibliography{./Literature}



\newpage

\appendix

\section{Definition for Precision, Accuracy, and Stability}
\label{Sec:PAS}

The use of the terms \textit{precision}, \textit{accuracy}, and \textit{stability} often lead to confusion among researchers.
A stringent and universal definition for these terms is difficult and their interpretation and application in practice might depend on the context and implicit assumptions. Also, different scientific fields adopt their own definitions or use other terms for the same concepts. We thus line out the definition adopted within this work.

\begin{figure*}
 \includegraphics[width=\linewidth]{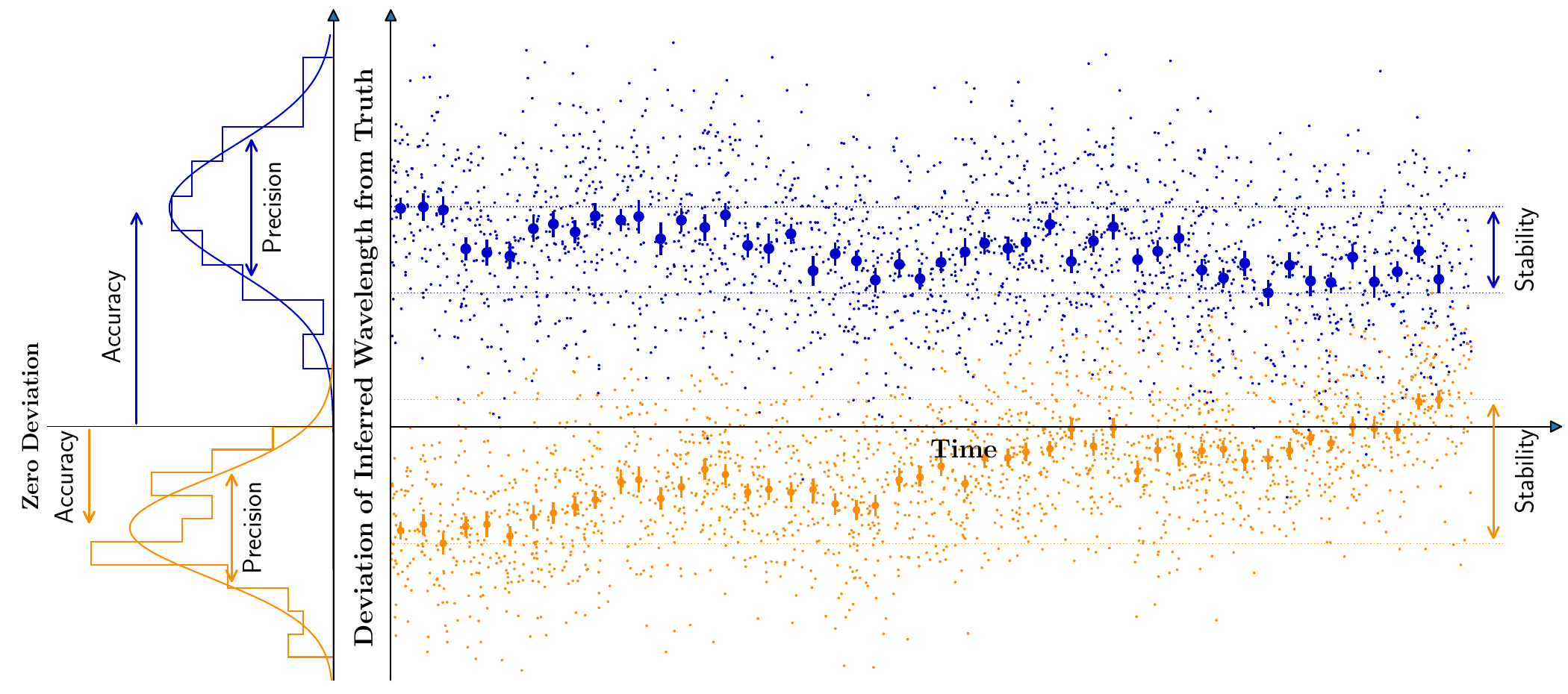}
 \caption{
 Visualization of the adopted concepts for \textit{precision}, \textit{accuracy}, and \textit{stability}, as defined in \citet{Martins2023}. Figure and caption are adopted from there with permission.
 The right panel shows a sequence of wavelength measurements (small dots) taken over a certain period of time. To visualize different types of effects, measurements for two spectral lines (blue and orange) are shown. Plotted for both of them is the deviation of the inferred value from the truth (horizontal black line). To aid the eye, measurements are grouped into bins (thick dots). The left panel shows histograms for the points falling into the first bin.
 \textit{Precision} is the scatter of formally identical measurements (due to stochastic effects), while \textit{accuracy} describes the offset or bias of this distribution from the true value.
 Over time, this offset will not be stable but be subject to drifts, which characterizes the \textit{stability}.
 }
 \label{Fig:Visualization_PrecisionAccuracyStability}
\end{figure*}

To specify the scientific requirements for the wavelength calibration of the ANDES spectrograph \citep{Marconi2022}, practical definitions for the terms \textit{precision}, \textit{accuracy}, and \textit{stability} have been worked out by the ANDES working group focused on fundamental physics and cosmology and were briefly described in \citet{Martins2023}. These definitions have been developed within the context of the fundamental physical constants project, i.e. searching for a possible change of the fine-structure constant \citep[e.g][]{Webb1999, Murphy2022b}, and the redshift drift experiment (a.k.a. the Sandage test, \citealt{Sandage1965}). The clear intention, however, was to make them as general as possible.
There is no claim by these working definitions to be universal or definitive and other suitable definitions might exists as well. However, since a considerable group of astronomers has agreed on a common definition and use of these terms, we also adopt this framework and apply it in the same, consistent way.
More essential than the actual terms, \textit{precision}, \textit{accuracy}, and \textit{stability}, are in fact concepts that stand behind thee terms and that are described in detail below and illustrated in Figure~\ref{Fig:Visualization_PrecisionAccuracyStability}.

For the measurement considered in the following, one might imagine the wavelength that one assigns, e.g. by performing a fit, to a single spectral feature (e.g. an absorption or emission line) in an observed spectrum.
However, the general concepts should apply as well to the result of any scientific measurement process.

\textbf{Precision:}
The \textit{precision} of a measurement quantifies the amount of purely stochastic scatter observed among equivalent repetitions of the same measurement process under fully identical conditions. In practice, this can be obtained by taking multiple measurements in quick succession, sufficiently fast so than no systematic effects come into play.
Alternatively, the magnitude of this scatter can also be estimated from first principles, e.g. from the amount of detector read-out noise and Poissonian photon noise.  Propagating these fundamental noise sources and taking into account additional well quantifiable, stochastic effects, e.g. related to the instrument, should as well yield a proper prediction for the precision of the measurement.
An important aspect is that the \textit{precision} can be improved by considering an ensemble of equivalent measurements (i.e. by binning or averaging), given that the scatter is indeed only driven by stochastic processes.

\textbf{Accuracy:}
While stochastic effects can be reduced by taking multiple measurements, there might remain an irreducible, systematic bias between the value inferred in the measurement process and the true value, which defines the \textit{accuracy}. Here, the comparison to a \textit{ground truth} value is a rather theoretical concept. In practice, such an absolute truth is often not available and has to be approximated.

For astronomical spectroscopy, every wavelength measurement is always relative, between the actual science observation and some calibration sources used to establish a wavelength solution for the spectrograph. The calibration sources are then connected to some more fundamental standard, i.e. their wavelengths are known in absolute terms or determined in laboratory experiments.
To achieve an accurate measurement in a science observation, one has to ensure that the calibration sources are accurately known by themselves and that instrument systematics affect science and calibration spectra in the same way.
To fully \textit{validate} the accuracy of a wavelength calibration, one would need a ground-trough source, on-sky, and with properties equivalent to a science target. At the level of interest, no astronomical source with so well-known properties exist.
The accuracy of a wavelength calibration can therefore not be verified against some on-sky reference, but only against other engineered validation sources within the telescope or instrument%
\footnote{There is not necessarily a technical difference between a \textit{calibration} and \textit{validation} source, but rather a conceptual one. The first one is used to establish a wavelength solution, the latter one to test and validate it. Within the context of our study, the LFC as well as the combined ThAr/FP system can both take the role of either calibration or validation source.}.
There could in principle be some systematics that affect both internal sources, but not the actual science observations. In that case, one would find consistency between calibration and validation source, but not achieve actual accuracy.
However, the more dissimilar various kinds of sources are (e.g. absorption lines vs. emission features, coherent vs. incoherent, narrow vs. resolved lines, ...), the more unlikely it becomes that some kind of systematic affects just calibration and validation source but not the actual science observation and thus the more representative their consistency becomes for the true accuracy of the measurement process.
In any case, the level of consistency observed among different calibration or validation sources sets a clear lower bound for the accuracy one can expect in an actual science observation. Another lower bound can be obtained from measuring the same quantity (i.e. consistency between sources) with different setups, e.g. in different spectral orders, fibers, or slices.

In summary, \textit{accuracy} can typically not be rigorously proven. One can only build confidence in the measurement process by eliminating obvious sources of inaccuracy, demonstrate consistency among various types of sources and over a variety of setups, and argue that additional systematics can be excluded beyond any reasonable doubt.

\textbf{Stability:}
The concepts lined out above describe the instantaneous properties of a measurement process. However, for many science cases, e.g. the search for extrasolar planets \citep[e.g][]{Mayor1995} or the redshift drift experiment \citep{Sandage1965, Liske2008, Cristiani2023}, the evolution of a measurement with time is of crucial importance. These studies basically do not care about the absolute value of the (wavelength) measurement, but only by how much and in which way it changes over some time. The figure of merit for such time-series observations is thus the \textit{stability} of the wavelength calibration, which describes by how much the result of a measurement changes with time due to adverse instrumental effects.
Here, numerous types of systematics can lead to instrumental drifts on various different timescales, e.g. diurnal or seasonal semi-periodic temperature variations, gradual effects like the shrinkage of the Zerodur spacer in FP etalons or the sagging of optical elements, probably enhanced by frequent micro-earthquakes, piece-wise monotonic effects like the aging of hollow-cathode lamps (until they get replaced), or disruptive events like large earthquakes, power-outages, or instrument interventions.

In Figure~\ref{Fig:Visualization_PrecisionAccuracyStability}, the stability is expressed as the time-variation of the systematic offset between measurements and true value, disregarding the purely stochastic scatter.
Conceptually, one might therefore call the \textit{stability} also \textit{stability of accuracy}.
However, from a practical point of view, this would be misleading and confusing. In practice, one would never first determine the absolute accuracy of a measurement process (which is as lined-out above rather difficult) and then asses its evolution.
Instead, the stability can be much better estimated by  directly comparing the measurements obtained at different times, irrespective of any offset from a ground-truth. This avoids dealing with numerous systematics that affect the accuracy, but remain constant in time. Some analysis methods even completely avoid assigning actual wavelengths to observed spectral features and only measure their drift with time \citep[e.g.][]{Bouchy2001, Dumusque2021}.
Thus, the \textit{stability} aspect of a measurement process can be treated completely detached from its \textit{accuracy}.
We also stress that the stability of astronomical wavelength measurements is often much better than their accuracy \citep[see e.g.][]{Cersullo2019}. For example, radial-velocity measurements of exoplanets, relying on the stability of the instrument and its calibration, are now routinely performed at the sub-m/s level \citep[e.g.][]{Faria2022}, but achieving a wavelength accuracy at the 10\,m/s level still poses significant challenges.

\section{Principle of measurement for the fine-structure constant}
\label{FineStructureConstant}

\begin{figure}
 \includegraphics[width=\linewidth]{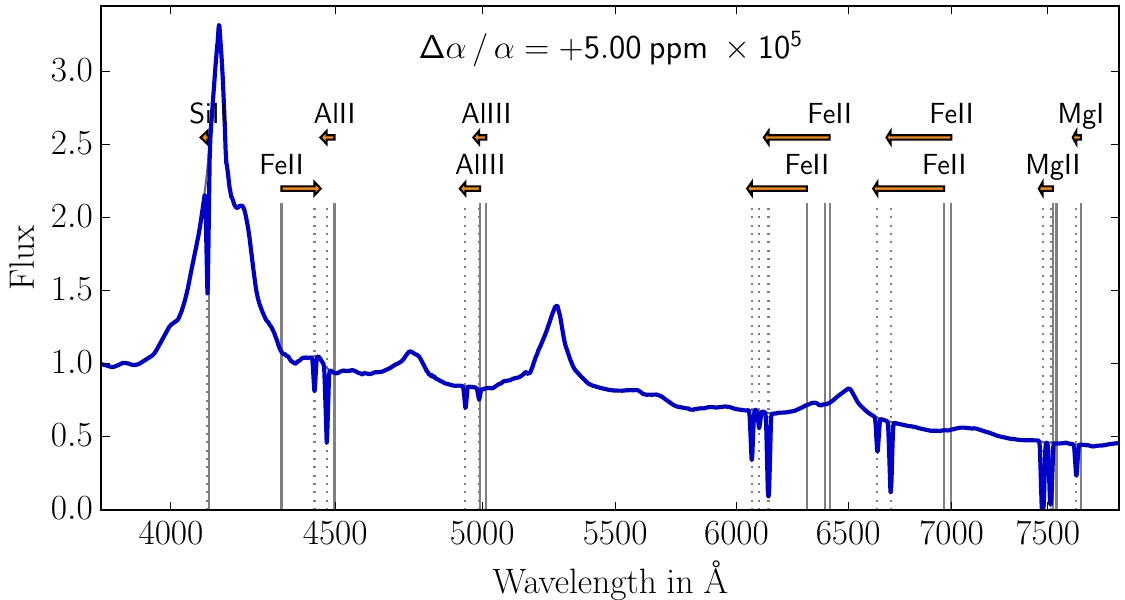}
 \caption{  Illustration of the general concept for measuring the fine-structure constant from quasar absorption systems, taken from \citet{Schmidt2021}. The figure shows an idealized quasar spectrum with a fiducial metal absorption systems at $z_\mathrm{abs}=1.7$.
  The assumed variation of $\Delta\alpha/\alpha$ by 5~parts-per-million (ppm) leads to differential shifts of the various absorption lines compared to their laboratory value. This is indicated by the arrows. For visibility, the magnitude of the shifts is exaggerated by $\times10^5$.
 }
 \label{Fig:Visualization_FineStructure}
\end{figure}

The general concept for measuring the fine-structure constant, $\alpha$, in quasar sightlines, following the same approach as e.g. \citet{Webb1999, King2012, Murphy2022}, and others, is visualized in Figure~\ref{Fig:Visualization_FineStructure}.
The fine-structure constant governs the coupling strength of electromagnetic interactions and therefore affects the energy levels of atomic transitions. A different value of $\alpha$ would lead to different absorption-line wavelengths. Fortunately, different transitions have a different sensitivity to $\alpha$ \citep[e.g.,][]{Dzuba1999a}. Thus, by observing multiple transitions with different sensitivity to $\alpha$ originating from the same absorption system and accurately measuring their wavelength allows to infer simultaneously the absorption redshift of the system and the value of the fine-structure constant. Each such measurement is in principle a single-shot experiment that yields the value of $\alpha$ at the place and time when the absorption happened. This can then be compared to the fine-structure constant measured on Earth or in other absorption systems.

The velocity shift for the absorption lines with the highest sensitivity (\ion{Fe}{ii}) is about $25\,\textup{m/s}$ for a $10^{-6}$ change of $\alpha$, while many other lines remain relatively unchanged (e.g. \ion{Mg}{ii}, \ion{Al}{iii}).
To avoid spurious detections of a change of $\alpha$ at the $10^{-6}$ level, the wavelength calibration thus has to be correct and in particular free of distortions at the $25\,\textup{m/s}$ level. A more rigorous assessment is given in \citet{Schmidt2021}.
This type of measurement therefore requires a great level of \textit{accuracy} of the wavelength solution, meaning that each individual absorption-line measurement actually needs to be correct.
The only exception is that a global offset of all inferred wavelengths by a fixed velocity would not change the inferred value of $\alpha$, since such a shift would be fully degenerate with the redshift of the absorption system.
However, apart from the barycentric correction, there is no effect that would just cause a constant velocity shift without introducing other distortions in the wavelength solution.
Therefore, strictly speaking, a measurement of the fine-structure constant requires accuracy of the wavelength solution up to a constant offset in velocity space.
However, for ease of wording, to avoid introducing wavelength calibration requirements that are correlated among different wavelengths, and because  there is basically no difference in practice, the requirement is typically phrased in the way that every single wavelength measurement needs to be accurate to a certain level.


\label{lastpage}
\end{document}